\documentclass[twocolumn, linenumbers]{aastex631}

\usepackage{siunitx}
\DeclareSIUnit{\belm}{Bm}
\DeclareSIUnit{\beli}{Bi}
\DeclareSIUnit{\dbeli}{dBi}
\newcommand{\SIadj}[2]{\SI[number-unit-product={\text{-}}]{#1}{#2}}
\usepackage[super]{nth}
\usepackage{xcolor}

\usepackage{amsmath}

\shortauthors{Dachlythra et al.}
\graphicspath{{./}{}}

\begin{document}
\nolinenumbers
\title{The Simons Observatory: Beam characterization for the Small Aperture Telescopes}

\author[0009-0006-7382-1434]{Nadia Dachlythra}
\affiliation{The Oskar Klein Centre, Department of Physics, Stockholm University, AlbaNova, SE-10691 Stockholm, Sweden}
\author[0000-0003-2856-2382]{Adriaan J. Duivenvoorden}
\affiliation{Center for Computational Astrophysics, Flatiron Institute, 162 5th Avenue, New York, NY, USA 10010}
\affiliation{Joseph Henry Laboratories of Physics, Jadwin Hall, Princeton University,  Princeton, NJ 08544, USA}
\author[0000-0003-1760-0355]{Jon E. Gudmundsson}
\affiliation{Science Institute, University of Iceland, 107 Reykjavik, Iceland}
\affiliation{The Oskar Klein Centre, Department of Physics, Stockholm University, AlbaNova, SE-10691 Stockholm, Sweden}
\author[0000-0002-2408-9201]{Matthew Hasselfield}
\affiliation{Center for Computational Astrophysics, Flatiron Institute, 162 5th Avenue, New York, NY, USA 10010}
\author[0000-0002-6362-6524]{Gabriele Coppi}
\affiliation{Department of Physics, University of Milano - Bicocca, Piazza della Scienza, 3 - 20126 Milano (MI), Italy}
\author[0000-0002-5736-5524]{Alexandre E. Adler}
\affiliation{The Oskar Klein Centre, Department of Physics, Stockholm University, AlbaNova, SE-10691 Stockholm, Sweden}
\author[0000-0002-4598-9719]{David Alonso}
\affiliation{Department of Physics, University of Oxford, Denys Wilkinson Building, Keble Road, Oxford OX1 3RH, UK}
\author[0000-0002-8132-4896]{Susanna Azzoni}
\affiliation{Department of Physics, University of Oxford, Denys Wilkinson Building, Keble Road, Oxford OX1 3RH, United Kingdom}
\affiliation{Kavli Institute for the Physics and Mathematics of the Universe (Kavli IPMU, WPI), UTIAS, The University of Tokyo, Kashiwa, Chiba 277-8583, Japan}
\author[0000-0001-6702-0450]{Grace E. Chesmore}
\affiliation{Department of Physics, University of Chicago, 5720 South Ellis Avenue, Chicago, IL 60637, USA}
\author[0000-0002-3255-4695]{Giulio Fabbian}
\affiliation{Center for Computational Astrophysics, Flatiron Institute, New York, NY 10010, USA}
\affiliation{School of Physics and Astronomy, Cardiff University, The Parade, Cardiff, CF24 3AA, UK}
\author[0000-0001-8159-8208]{Ken Ganga}
\affiliation{Universit\'e Paris Cité, CNRS, Astroparticule et Cosmologie, F-75013 Paris, France}
\author{Remington G. Gerras}
\affiliation{Department of Physics, University of Southern California, Los Angeles, CA 90089, USA}
\author[0000-0003-2086-1759]{Andrew H. Jaffe}
\affiliation{Astrophysics Group and Imperial Centre for Inference and Cosmology, Department of Physics, Imperial
College London, Blackett Laboratory, Prince Consort Road, London SW7 2AZ, United Kingdom}
\author[0000-0002-6898-8938]{Bradley R. Johnson}
\affiliation{University of Virginia, Department of Astronomy, Charlottesville, VA 22904}
\author[0000-0003-3118-5514]{Brian Keating}
\affiliation{Department of Physics, University of California, San Diego, La Jolla, CA 92093, USA}
\author[0000-0001-5748-5182]{Reijo Keskitalo}
\affiliation{Computational Cosmology Center, Lawrence Berkeley National Laboratory, Berkeley, CA 94720, USA}
\affiliation{Department of Physics, University of California, Berkeley, CA 94720, USA}
\author[0000-0003-3510-7134]{Theodore S. Kisner}
\affiliation{Computational Cosmology Center, Lawrence Berkeley National Laboratory, Berkeley, CA 94720, USA}
\affiliation{Department of Physics, University of California, Berkeley, CA 94720, USA}
\author[0000-0002-5501-8449]{Nicoletta Krachmalnicoff}
\affiliation{International School for Advanced Studies (SISSA), Via Bonomea 265, I-34136 Trieste, Italy}
\affiliation{National Institute for Nuclear Physics (INFN) – Sezione di Trieste, Via Valerio 2, I-34127 Trieste, Italy}
\affiliation{Institute for Fundamental Physics of the Universe (IFPU), Via Beirut 2, I-34151 Grignano (TS), Italy}
\author[0000-0002-3800-5558]{Marius Lungu}
\affiliation{New York, NY, USA}
\author[0000-0003-0041-6447]{Frederick Matsuda}
\affiliation{Japan Aerospace Exploration Agency (JAXA), Institute of Space and Astronautical Science (ISAS), Sagamihara, Kanagawa 252-5210, Japan}
\author[0000-0002-4478-7111]{Sigurd Naess}
\affiliation{Institute of theoretical astrophysics, University of Oslo, Norway}
\author[0000-0002-9828-3525]{Lyman Page}
\affiliation{Joseph Henry Laboratories of Physics, Jadwin Hall, Princeton University, Princeton, NJ 08544, USA}
\author{Roberto Puddu}
\affiliation{Instituto de Astrofísica and Centro de Astro-Ingeniería, Facultad de Física, Pontificia Universidad Católica de Chile, Macul, Santiago, Chile}
\author[0000-0002-0689-4290]{Giuseppe Puglisi}
\affiliation{Università di Catania, Dipartimento di Fisica e Astronomia, Sezione Astrofisica, Via S. Sofia 78, 95123 Catania, Italy}
\affiliation{INAF - Osservatorio Astrofisico di Catania, via S. Sofia 78, 95123 Catania, Italy}
\author[0000-0001-9221-7802]{Sara M. Simon}
\affiliation{Fermi National Accelerator Laboratory, Batavia, IL, 60510, USA}
\author{Grant Teply}
\affiliation{Department of Physics, University of California, San Diego, La Jolla, CA 92093, USA}
\author{Tran Tsan}
\affiliation{Department of Physics, University of California, San Diego, La Jolla, CA 92093, USA}
\author[0000-0002-7567-4451]{Edward J. Wollack}
\affiliation{NASA Goddard Space Flight Center, 8800 Greenbelt Road, Greenbelt, MD, 20771, USA}
\author{Kevin Wolz}
\affiliation{International School for Advanced Studies (SISSA), Via Bonomea 265, I-34136 Trieste, Italy}
\affiliation{National Institute for Nuclear Physics (INFN) – Sezione di Trieste, Via Valerio 2, I-34127 Trieste, Italy}
\author[0000-0001-5112-2567]{Zhilei Xu}
\affiliation{MIT Kavli Institute, Massachusetts Institute of Technology, 77 Massachusetts Avenue, Cambridge, MA 02139, USA}


\begin{abstract}

We use time-domain simulations of Jupiter observations to test and develop a beam reconstruction pipeline for the Simons Observatory Small Aperture Telescopes. The method relies on a map maker that estimates and subtracts correlated atmospheric noise and a beam fitting code designed to compensate for the bias caused by the map maker. We test our reconstruction performance for four different frequency bands against various algorithmic parameters, atmospheric conditions and input beams. We additionally show the reconstruction quality as function of the number of available observations and investigate how different calibration strategies affect the beam uncertainty. For all of the cases considered, we find good agreement between the fitted results and the input beam model within a $\sim$ 1.5$\%$ error for a multipole range $\ell$ = 30 -- 700 and a $\sim$ 0.5$\%$ error for a multipole range $\ell$ = 50 -- 200. We conclude by using a harmonic-domain component separation algorithm to verify that the beam reconstruction errors and biases observed in our analysis do not significantly bias the Simons Observatory $r$-measurement.

\end{abstract}


\section{Introduction} \label{sec:intro}
The temperature anisotropy of the Cosmic Microwave Background (CMB) has been mapped across a wide range of angular scales \citep[see e.g.,][]{Bennett_2013, planck_overview_2020}. Information in the polarization anisotropies, which are significantly weaker, has yet to be characterized as extensively. Continued measurements of the CMB polarization will help break the degeneracy between various cosmological parameters and provide an additional probe into the cosmic inflation paradigm. For the latter case, the community is focusing on measuring the power of the parity-odd polarization component, the so-called B-mode polarization, on degree scales and larger, that could be directly sourced by a primordial background of stochastic gravitational waves, a key prediction of some inflationary scenarios \citep[see e.g.,][]{new_physics_from_CMBpol}. It is common practice to quantify the amplitude of the primordial B-mode polarization in terms of the tensor-to-scalar ratio, \textit{r}.

The Simons Observatory (SO) Small Aperture Telescopes (SATs) aim to constrain the tensor-to-scalar ratio with unprecedented sensitivity, targeting a statistical error of $\sigma (r)$ = 0.003 \citep{Ade2019} or better. In order to do so, a collection of \SIadj{42}{\centi\meter} aperture SATs will observe the CMB temperature and polarization from a \SI{5200}{\meter} altitude at the Atacama desert in Chile. Observations will be done in six frequency bands to allow mitigation of Galactic foregrounds \citep[see e.g.,][]{Krachmalnicoff_2016}. The tensor-to-scalar ratio constraint is an ambitious goal that calls for a comprehensive understanding of our telescopes' performance.  

Improper beam modeling can significantly bias the telescope's science goals. A small beam reconstruction error between different frequency bands is important for the success of foreground component separation analyses. The B-modes from the polarized Galactic foregrounds are much stronger than the primordial signal we are seeking, and thus, a slightly biased estimation of the amplitude of the foregrounds due to calibration mismatch can lead to an important bias on the tensor-to-scalar ratio. Furthermore, recovering the beam transfer function with a small error also facilitates the calibration against Planck data. This calibration will happen at intermediate angular scales where the smoothing effect of the beam is important. Biased estimates of the beams will again lead to relative biases between the frequency bands, which may significantly bias the inference of the primordial B-mode amplitude.

The main beam systematics represents only a small fraction of the long list of optical systematics that can impact cosmological analysis of data from small aperture CMB telescopes. These include beam asymmetries of various types; beam sidelobes; polarization angle errors; internal reflection causing so-called ghosting; pointing errors and half-wave-plate-related systematics, including spurious scan-synchronous effects. For efforts related to constraining amplitude of primordial B-mode polarization, it is perceivable that all of the effects listed above could be non-negligible. Many of these effects are discussed in the following publications: \cite{Shimon_2008, Fraisse_2013, planck_toi_and_beams_2015, salatino2018studies, BK2019, Xu_2020, Abitbol_2021, beamconv_hwp_2021}. The accurate determination of azimuthally averaged Stokes I beam profiles for Simons Observatory Small Aperture Telescopes represents a necessary, but not a sufficient requirement for accurate constraints of the amplitude of primordial B-mode polarization

Beam calibration techniques for CMB telescopes have been investigated in a number of publications \citep[see e.g.,][]{bicep2_beamcal2010, Keating_2013, Hasselfield_2013, planck_toi_and_beams_2015, BK2019, Lungu_2022}. This paper adds to the existing literature by investigating the observational requirements and capabilities for beam reconstruction for the SO SATs. Although optical design software can be used to predict the far-field beam response, the final beam model used for science analysis will rely heavily on planet observations that are made through fluctuating atmosphere. It is therefore important to develop algorithms that accurately capture the details of such observations. This involves creating simulations that include realistic detector noise and atmospheric emission and using those to show how our beam reconstruction depends on observation time, properties of atmospheric emission, and low-frequency thermal variations in our instrument. 

The paper is structured as follows: Section~\ref{sec:instrument} describes the SAT instrument design and physical optics models which will be used throughout the paper. 

Section~\ref{sec:pipeline} summarizes the simulation pipeline. 
Details of analysis methods are described explicitly in Section \ref{sec:map_d_fit} with key results summarized in Section \ref{sec:results}. Section \ref{sec:conclusion} offers  conclusions and discussion.

\section{Instrument design and beam modeling} 
\label{sec:instrument}

The Small Aperture Telescopes use a three-lens cryogenically cooled silicon refractor design \citep{sat_optics_design}. The optics have a \SIadj{42}{\centi\meter} diameter aperture and support a wide field-of-view (\SI{35}{\degree}) \citep{galitzki2018simons}. The cryo-mechanical and optical design is described in \cite{Ali_2020}. Each SAT can support up to approximately 10,000 dichroic detectors occupying in total 7 hexagonal, \SIadj{150}{\milli\meter} diameter, silicon wafers forming a (also hexagonal) focal plane with a radius of approximately \SI{17.5}{\centi\meter} (see Figure 1 of \cite{galitzki2018simons}). The detectors are cooled to \SI{100}{\milli\kelvin} while the lenses and aperture stop are cooled to \SI{1}{\kelvin}. The dichroic detectors operate at two Low-Frequency (LF), two Mid-Frequency (MF), and two Ultra-High-Frequency bands (UHF), centred near 27 and 39, 93 and 145, and 225 and \SI{280}{\giga\hertz}.

The SATs are equipped with a cryogenically cooled half-wave plate (HWP) mounted skywards of the optics. The spinning HWP (at \SI{2}{\hertz}) modulates the linearly polarized component of the sky in a controlled fashion. Any unpolarized signal from the sky and atmosphere is left unmodulated, which  suppresses temperature-to-polarization (T-to-P) leakage and allows for a clean measurement of the polarized sky signal.

More details about the HWP design and related studies for Simons Observatory can be found in \cite{Hill_2018, salatino2018studies}. The telescopes are externally baffled to suppress signal from the ground and nearby mountains. Specifically, each SAT telescope has a free-standing ground shield, a nominally reflective co-moving shield and a nominally absorptive forebaffle. Diffraction caused by baffling elements can potentially create polarized beam sidelobes that couple to both the ground and the Galaxy; modelling of the effect for a shielded refractor has been investigated in \cite{Adler_2020}.

\begin{figure}
    \centering    \includegraphics[width=\linewidth]{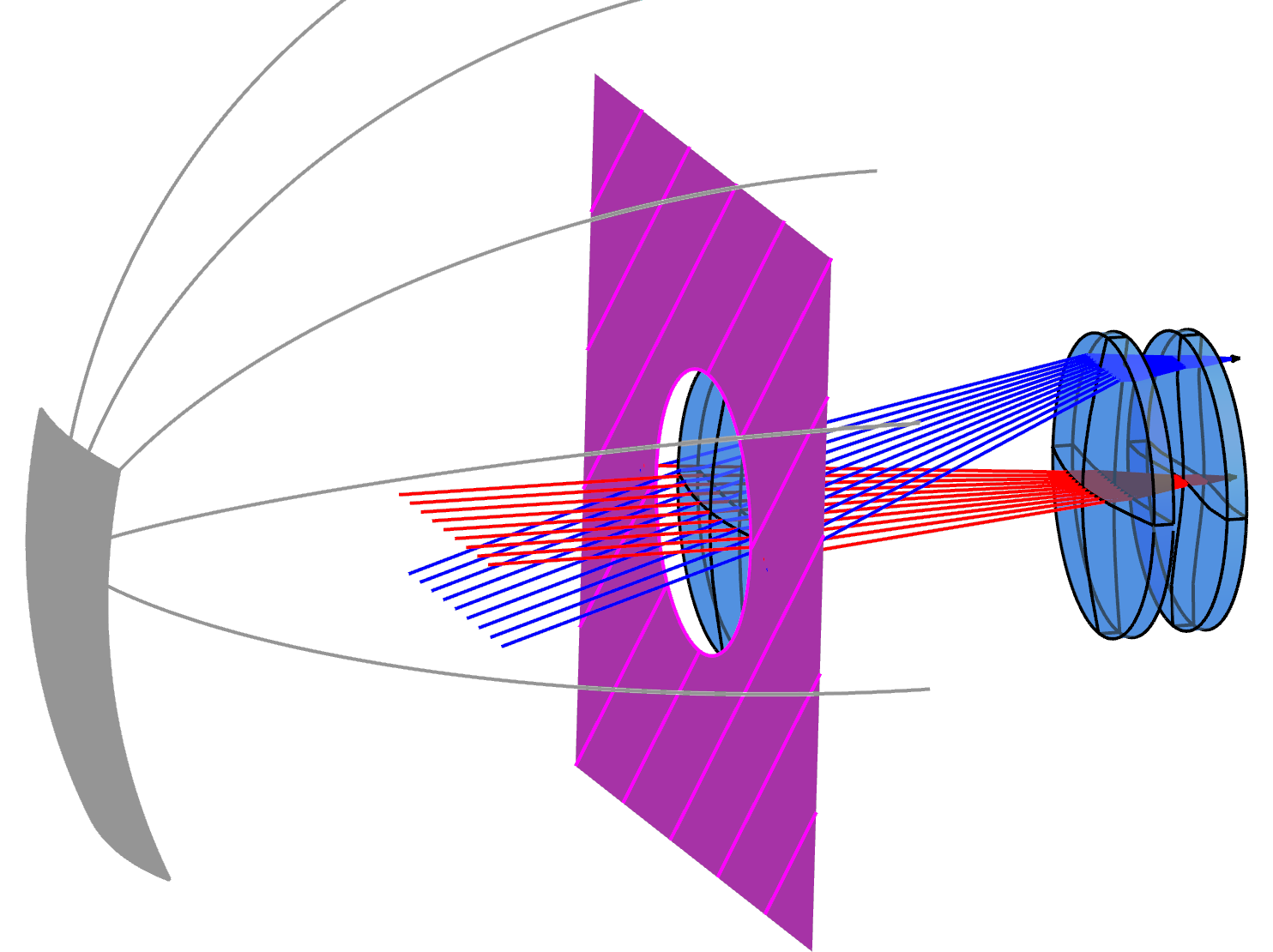}
    \caption{\label{fig:GRASP_setup} The 3-lens SO SAT refracting telescope design which was implemented in $\texttt{TICRA TOOLS}$ for the production of far-field beam maps for the SATs. In this setup, the light rays (in time-reversed simulations) travel from a centre (red lines) and edge pixel (blue lines) of the focal plane through the three lenses (blue) and the aperture stop (purple surface) towards the far-field (the grey surface, not to scale) where the output beam is tabulated. The distance from the focal plane to the sky side of the primary lens is approximately \SI{81}{\centi\meter}. The diameter of the three silicon lenses is about \SI{45}{\centi\meter}.}
\end{figure}

The beam models for the SO telescopes are generated using Ticra Tools\footnote{TICRA, Landemærket 29, Copenhagen, Denmark (https://www.ticra.com)} (formerly GRASP), proprietary software based on physical optics and the physical theory of diffraction. With Ticra Tools, we simulate various optical components such as lenses, antennas, feed-horns and stops allowing us to capture critical features of the SO SAT design. For this analysis, we simulate the 2D far-field co- and cross-polar beam maps for pixels at various locations on the focal plane. Figure~\ref{fig:GRASP_setup} shows a representative telescope configuration as set up in Ticra Tools.

\begin{figure}[t!]
    \centering
    \includegraphics[width=\linewidth]{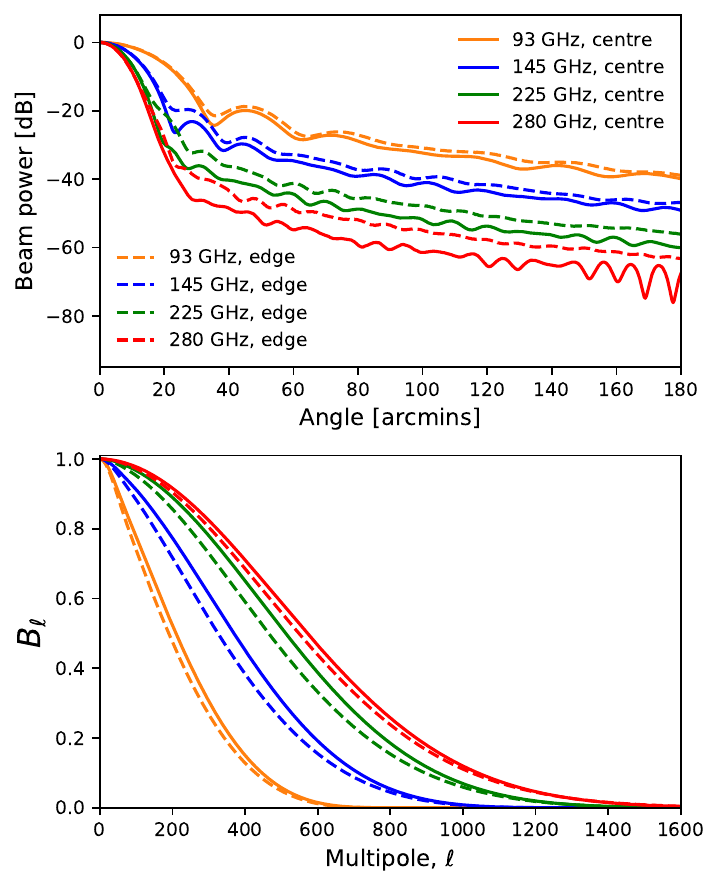}
    \caption{\label{fig:input_beam_profiles}Top: Beam profiles of the band-averaged input beam models for the four SO SATs frequency bands discussed in this paper. Solid and dashed lines show the cases for a detector placed on the center (0 cm) and the edge (18 cm) of the focal plane, respectively. The beam profiles are computed by radially binning the 2D band-averaged beam maps. Bottom: Transfer functions for the beam models whose profiles are presented in the top plot.}
\end{figure}

We simulate beam maps for four frequency bands centred on 93, 145, 225 and \SI{280}{\giga\hertz}. We will be referring to those four frequency bands as Mid-Frequency 1 (MF1), Mid-Frequency 2 (MF2), Ultra-High-Frequency 1 (UHF1) and Ultra-High-Frequency 2 (UHF2). For each, we make band-integrated maps from five single-frequency simulations over a 20$\%$ bandwidth around the centre frequency. For the nominal input models we use throughout the paper, we assume a top-hat spectral response function, although different weighting schemes can be implemented trivially. A class of potential input beam models for the SATs assuming non-uniform passband and different types of frequency scaling are shown in Appendices \ref{appendix:var_in_beam} and \ref{appendix:var_in_sed} for reference.

Figure \ref{fig:input_beam_profiles} shows the SAT beam profiles (top) and corresponding transfer functions (bottom) for the band-averaged simulations, assuming a pixel at the centre (solid lines) and edge (dashed lines) of the focal plane, respectively. From the top panel of the figure, we see that the beam profiles are approximately Gaussian in the centre with a sidelobe at larger angles, where the beam power roughly drops as the inverse cube of the angle. The edge pixel is located \SI{18}{cm} from the centre of the focal plane, corresponding to a beam centroid that is shifted by $\sim$ \ang{17.5} relative to the telescope boresight. The centre and edge pixel beam models shown in this figure will be assigned to all detectors of the centre and one of the edge wafers correspondingly when simulating timestreams.

 The simulated far-field maps correspond to the co-polar component of the beam. The cross-polar component is small in amplitude and should be studied together with the instrument's HWP performance. As planets are mostly unpolarized, observed polarization would be the result of T-to-P leakage, which the HWP failed to prevent (see Section \textcolor{blue}{5} of \cite{Lungu_2022}). This part of the analysis is left for future work. 

 The SAT beams are treated as azimuthally symmetric throughout the paper, a choice that is strongly motivated by the GRASP simulations. As we will be scanning the sky both when rising and setting, the cross-linking in temperature maps will, furthermore, symmetrise the beams. Any T-to-P leakage caused by remaining beam asymmetry is expected to be suppressed by the spinning HWP \citep{salatino2018studies}.

\begin{center}
\begin{table}[h!]
\centering
\begin{tabular}{ c c c c} 
 \hline
  Frequency-band & FWHM & Ellipticity & Solid angle \\ 
&[arcmins] & $\epsilon$ & [$10^{-6} $ sr] \\
 \hline
 MF1  (93 GHz) & 27.4 & 0.030 & 78.9 \\ 
 MF2 (145 GHz) & 17.6 & 0.036 & 30.5 \\ 
 UHF1 (225 GHz) & 13.5 & 0.046 & 17.3 \\ 
 UHF2 (280 GHz) & 12.1 & 0.045 & 13.6 \\ 

 \hline
\end{tabular}
\caption{\label{tab:beam_params} Best-fit beam size (FWHM), ellipticity and solid angle per frequency band for the simulated SAT beams, assuming a pixel placed at the focal plane centre. The forward gain derived from the total beam solid angle is 52.0, 56.2, 58.6 and \SI{59.7}{\deci\beli} for MF1, MF2, UHF1, and UHF2, respectively.}
\end{table}
\end{center}
Table \ref{tab:beam_params} provides an overview of the simulated beams in terms of Full-Width-Half-Maximum (FWHM), solid angle, and the best-fit value for the beam ellipticity, defined as:
\begin{equation}
\epsilon = \frac{\theta_{\mathrm{maj}}-\theta_{\mathrm{min}}}{\theta_{\mathrm{maj}}+\theta_{\mathrm{min}}},
\end{equation} 
where $\theta_\mathrm{maj}$ and $\theta_\mathrm{min}$ are the FWHM of the beam's major and minor axis, respectively. 
The values in Table \ref{tab:beam_params} are determined by fitting 2D elliptical Gaussians to the beam maps and apply to a pixel placed on the center of the focal plane. Our estimates suggest that the beam FWHM of a pixel located at the edge of the focal plane will differ by about 1-2~\% from its centre-pixel value while the beam ellipticity may change by a factor of $\sim 50\%$ from the center to the edge of the focal plane. The results presented in this paper, however, are shown to be largely insensitive to the predicted variation in beam ellipticity across the focal plane (see Section \ref{sec:detector_position}).

\section{Simulation pipeline}
In this section, we discuss potential calibration sources, and describe the software and scan strategy employed to simulate the time domain data from observations of a bright point source, given the beam model discussed in the previous section.
\label{sec:pipeline}

\subsection{Candidate sources}
\label{sec:inputs}

Due to beam dilution, only a handful of natural point sources exist that are bright enough to calibrate the SAT beams. Of these, Jupiter is the brightest \citep{calibration_wmap_2011, planck_planetflux_2017} and most suitable for SAT calibrations from the Atacama desert. We focus on the characterization of the instrument's response to an unpolarized source as a baseline. The polarization response, which additionally requires a measurement of instrument's polarization angle and cross-polar beam components is left for future work. 

Artificial calibration sources have the potential to overcome the limitations of the astrophysical sources. At the moment  sources mounted on tall structures have been successfully used for beam calibration \citep{BK2019}. For the future balloons \citep{Masi_2006}, drones \citep{9411058}, and even satellites \citep{Johnson2015} are being considered. The use of drones for calibration purposes is the subject of multiple active studies \citep{Nati_2017}. Calibration sources mounted on drones can be tuned in brightness, frequency, and cadence in order to meet the calibration requirements of different instruments. Additionally, these sources can be equipped with a polarizing wire grid which facilitates the calibration of polarization intensity and angle \citep{2020SPIE11453E..2PD}. This last aspect is particularly important as there are very few polarized astrophysical sources that are bright enough to calibrate the polarization response of the SO SATs (the highest upper limit of the planets polarization fraction, $p_{\mathrm{frac}}$, is assigned to Uranus and corresponds to $p_{\mathrm{frac}}<3.6 \%$ at \SI{100}{\giga\hertz} within 95$\%$ confidence limits \citep{planck_planetflux_2017}).  

There is a trade-off between calibrating with astrophysical and man-made sources. In the case of drones, flight endurance, especially in the thin atmosphere at high altitudes, is one of the main obstacles to successful calibration campaigns for experiments such as the Simons Observatory. After recent on-site testing, the maximum flight time for the drone has been established to be $\sim$ \SI{12}{\minute} \citep{Coppi:2022qjs}. For this reason, the nominal calibration strategy relies on the planets.

\begin{figure}
    
    \centering  \includegraphics[width=\linewidth]{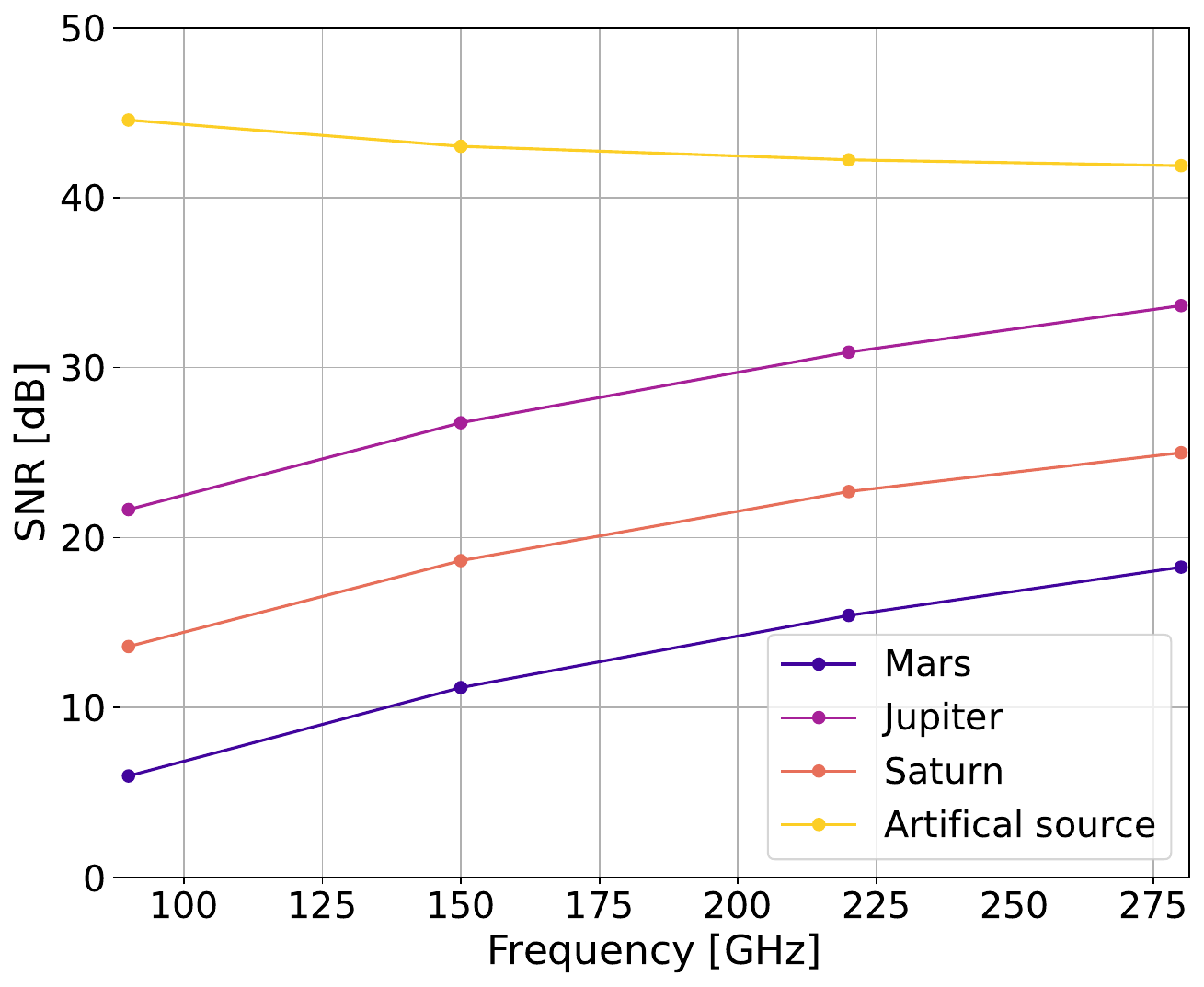}   \caption{\label{fig:planet_vs_drone} Comparison between the expected SNR when observing an artificial source mounted on a drone versus planet observations, for a single detector as a function of frequency. For all cases included in this plot, the noise has been considered to be in the range of $20-\SI{90}{\atto\watt / \sqrt{\hertz}}$, depending on the band, with a $\SI{1}{\second}$ integration time and $20\%$ bandwidth. The artificial source is assumed to be at a $\SIadj{500}{\meter}$ distance from the telescope (with a power output of $\SI{-18}{\deci\belm}$ at all frequencies and an antenna gain of $\SI{6.5}{\deci\beli}$).}
\end{figure}

Figure \ref{fig:planet_vs_drone} shows a comparison of the Signal-to-Noise Ratio (SNR) we can achieve when observing some of the brightest planets and when observing a source mounted on a drone as a function of frequency. The SNR estimation relies on the source's power and takes into account the Net-Equivalent-Power (NEP) and optical efficiency of the telescope while assuming the same integration time per pixel for all sources. When using the drone, we can expect a significantly higher SNR, which scales more smoothly with frequency compared to the various planets' cases. Note that the planets' brightness values are calculated as a function of their average estimated distance from the Earth over the year 2023. The exact calculations leading to the SNR values of Figure \ref{fig:planet_vs_drone} are described in Appendix \ref{appendix:snr_calculations}.

\subsection{The \texttt{TOAST} and \texttt{sotodlib} software}
\label{sec:toast_sotodlib}

The simulated time-ordered data of the Jupiter observations are generated with the help of the Time-Ordered Astrophysics Scalable Tools (\texttt{TOAST}\footnote{\textcolor{blue}{https://github.com/hpc4cmb/toast}}) library and the \texttt{sotodlib}\footnote{\textcolor{blue}{https://github.com/simonsobs/sotodlib}} library, that interfaces with TOAST and provides experiment configuration files that are specific to SO.

The \texttt{TOAST} software was developed for simulating, gathering and analyzing telescope time-ordered data. It is open-source software which is used in the framework of many current and next-generation CMB telescopes like LiteBIRD, Simons Array, Simons Observatory and CMB-S4. \texttt{TOAST} has been used in a recent study of instrumental systematics for experiments aiming to observe the CMB polarization \citep{TOAST_systematics_2021}.

The software can generate instrumental noise, atmospheric noise and scan-synchronous signals from ground pickup as well as simulate the effect of a HWP. The code is end-to-end parallelized and optimized for low memory consumption, which facilitates its use with workload managers on large servers. \texttt{TOAST} allows one to simulate sky observations for different scan strategies of tunable parameters (scanning speed, observing time and sky patch among others), and implement various sky models. 

The \texttt{TOAST} simulator module creates a focal plane configuration based on specified hardware parameters and samples the beam over a sky patch specified by the scheduler.
 
To simulate the atmosphere with \texttt{TOAST}, an additional file containing weather parameters is required. The corresponding module creates an atmospheric volume that moves with constant wind speed, set by the user or randomly drawn, and observed by individual detectors on the focal plane. This part of the code needs a set of input parameters, for example, for the detector gain, the field of view, and 
the center and width of the dissipation and injection scales of the Kolmogorov turbulence, describing atmospheric fluctuations \citep{Kolmogorov, Errard_2015}. In this work we limit ourselves to an atmospheric model based purely on water vapor, as experience proves it is the dominant disruption for CMB experiments \citep{Morris_2022}. Other potential absorbers such as clouds, ice crystals and oxygen are left for future work.

\subsection{Scan strategy and noise parameters}\label{sec:scan_strategy}

\begin{figure}
    \centering
    \includegraphics[width=\linewidth]{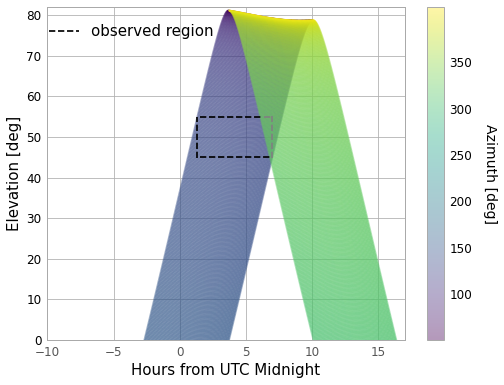}
    \caption{\label{fig:sso_ss} Jupiter's daily trajectory over the three months we are simulating (from June to August of 2021). Each of these curves shows the planet's elevation as a function of Universal Time Coordinated (UTC) time and is color-coded according to its azimuth value. The elevation of the observing region (indicated by the black dashed line) lies between 45 and \SI{55}{\degree} and corresponds to $\sim$ 1 hour of scanning.}
\end{figure}

For the Jupiter observations, we simulate Constant Elevation Scans (CES) to avoid systematic effects arising from varying the telescope's elevation, and we set a maximum allowed observing time of an hour. Given an observing site, time and target, \texttt{TOAST}  simulates observing schedules that conform to observing constraints such as elevation and boresight distance to Sun and the Moon. It uses the \texttt{PyEphem}\footnote{\textcolor{blue}{https://rhodesmill.org/pyephem/}} software to predict the target's location with respect to the observatory. We run the scheduler with an allowed elevation range of [\SI{45}{\degree}, \SI{55}{\degree}] and Sun/Moon avoidance radii set to \SI{45}{\degree}. The sampling rate and scanning speed are set to \SI{200}{\hertz} and \SI{1.5}{\degree\per\second}, respectively, at all times. The chosen maximum duration for a single CES facilitates tuning and calibrating the instrument; however, it takes Jupiter approximately 55 minutes to pass through the observing patch. After scheduling a single CES that follows these requirements, the scheduler moves on to the next day when Jupiter is available for observation. The simulated scans consider a single wafer (equipped with 860 detectors) each time instead of the full focal plane and cover an azimuth range of $\sim$ \SI{20}{\degree}. In practice, we will measure Jupiter with the HWP continuously spinning but do not model it in the simulation.
The signal amplitude from the sources is orders of magnitude greater than the polarized signal from the microwave sky and thus we neglect this effect in the current simulations.
Figure \ref{fig:sso_ss} shows the daily trajectory of Jupiter over the full months of June, July, and August of the year 2021 which we choose to simulate in terms of the planet's elevation as a function of time. Each curve is color-coded according to Jupiter's azimuth value. The 2D interval defined by the black dashed line refers to the observed region.

It should be noted that the chosen azimuth and elevation ranges of the observing patch are not yet set in stone, and we expect them to evolve as we move closer to the telescope’s deployment. This is because the choice of these parameters is primarily motivated by the source's availability during the observing period. The lower limit of the simulated elevation range is slightly lower than the most recent specifications, which set the nominal scanning elevation at \SI{50}{\degree} degrees. 
As atmospheric loading worsens with decreasing elevation, our chosen strategy should be considered as a slightly pessimistic case although we do not anticipate increasing the elevation range by \SI{5}{\degree} to improve our results significantly. 

For the chosen observation period and scan strategy described above, we accumulated $\sim$50 hours of Jupiter simulations in total, as some days the planet was not observable due to observing elevation and solar/lunar avoidance constraints. These simulations include atmospheric emission of both fixed and fluctuating weather parameters (see discussion in Section \ref{weather_conditions}). For the scope of this project, we consider the atmosphere to be unpolarized even though it can intermittently carry a non-negligible polarization fraction \citep{atm_pol_2019}. Furthermore, the simulations include realistic white and red detector noise expressed in Noise-Equivalent-Temperature (NET). A detailed description of the noise model and parameters for the SATs can be found in Table \textcolor{blue}{1} of \cite{Ade2019}. For the planet temperature calculation, we rely on the thermodynamic temperatures listed in Table \textcolor{blue}{4} of \cite{planck_planetflux_2017}. The retrieved temperature values are then interpolated to the frequency range we wish to simulate and converted to $T_\mathrm{CMB}$ units which express temperature in terms of the offset from the mean CMB temperature value.

\section{Analysis pipeline} \label{sec:map_d_fit}
We describe the map-maker applied to the Jupiter simulations and the associated atmospheric noise mitigation techniques. We offer insights into the different parameters of the algorithm and summarize the method used to fit the radial beam profiles and transfer functions.

Both the map-making and beam fitting are based on methods developed for ACT \citep{Hasselfield_2013, Lungu_2022}. The main new method development for this paper is the noise mitigation approach described in Section~\ref{sec:map_making}. Although the method is qualitatively similar to those used in the above references papers, our new implementation allows for more fine-grained tuning of the noise subtraction compared to the ACT implementations. Additionally, the previous implementations have not been described in full detail in the literature, so we provide a detailed description here. The size of the telescope's field-of-view compared to the angular scale of the atmospheric fluctuations largely determines the effectiveness of the noise subtraction. A larger field-of-view will lower the effectiveness. 
We describe in detail how our implementation is tuned to take into account the large field-of-view of the SO SAT compared to ACT and demonstrate that the noise subtraction is still sufficiently effective. 


\subsection{Map-making and low-level processing}\label{sec:map_making}

The simulation of planet observations starts with the generation of signal timestreams by using the \texttt{TOAST} and \texttt{sotodlib} software. The timestreams include white and correlated noise and are produced for all the detectors of each of the seven SAT focal plane wafers, as described in Section \ref{sec:scan_strategy}. However, the results shown in this paper only concern the center and one of the edge wafers. We gather all the timestreams simulated this way and construct $\SI{10}{\degree}\times \SI{10}{\degree}$ maps around the planet. 

We do not use a standard maximum-likelihood (ML) mapmaker for this analysis. While these are, in principle, unbiased and optimal and would therefore appear to be an ideal choice for measuring the instrument beam, in practice, they are only unbiased if the data precisely follow the fitted model. In reality, this is never the case. In this case, unmodeled gain and pointing fluctuations mean the observed signal is time-dependent in a way that a static image of the sky cannot capture. The result of such model errors is a bias that is typically at the sub-percent level but is non-local and is spread out by roughly a noise correlation length. This bias is large enough to completely overwhelm the fainter wings of the beam profile \citep{Naess_2019_x}.

To avoid this bias, we use a specialized filter-and-bin mapmaker that uses our knowledge of the planet's position to build a filter that removes as much of the atmosphere as possible while leaving the planet's signal almost untouched (see \cite{Hasselfield_2013}, \cite{Choi_2020}). In particular, if we assume that the planet's signal is entirely contained inside a mask with radius $\theta_\mathrm{mask}$ around its location, and that the noise is correlated with covariance matrix $\mathbf{C_n}$, then we can use all the data outside of the mask to make a prediction about the noise inside the mask and subtract it \citep{Lungu_2022}.

\begin{equation}
 \mathbf{d'} = \mathbf{d} - \underset{\mathbf{n}}{\arg\max} \: P(\mathcal{\textbf{n}} | \mathbf{d}_{\theta > \theta_{\mathrm{mask}}}, \mathbf{C_{n}}).  
 \label{eq:data_vector}
\end{equation}
Here $\mathbf{d'}$ and $\mathbf{d}$ are the raw and clean data vectors, respectively, \textbf{n} is the noise vector, and $\textbf{d}_{\theta>\theta_\mathrm{mask}}$ are the data outside of the mask. To the extent that all the signal is contained inside the mask, this subtraction will not introduce a bias. In practice, a small part of the signal will extend outside $\theta_\mathrm{mask}$, and there will be a trade-off between bias and noise subtraction (see discussion below).

For computational efficiency and to keep the implementation simple, we do not maximize $P$ in Equation \ref{eq:data_vector}, but instead approximate it using the $N_\mathrm{modes}$ strongest principal components of a copy of $\mathbf{d}$ where the area inside the mask has been filled in using polynomial interpolation. These principal components are then subtracted from the original $\mathbf{d}$ to form the cleaned $\mathbf{d'}$. Effectively, we're using the detectors outside the mask at any given moment to predict what correlated noise the detectors inside the mask should be seeing, and subtracting that. This approximation ignores the temporal correlations of the noise, but seems to perform sufficiently well for our configuration.

After this cleaning, we assume that any remaining noise in $\mathbf{d'}$ is uncorrelated, and can be mapped using a simple inverse-variance-weighted binned mapmaker. Note that the resulting map is only low-bias inside the masked region $\theta < \theta_\mathrm{mask}$. Any data outside $\theta_\mathrm{mask}$ are effectively highpass filtered due to the noise subtraction and, thus, heavily biased.

The effectiveness of this method depends strongly on the signal being compact (so one can use a small $\theta_\mathrm{mask}$) compared to the correlation length of the noise. It is therefore best suited for high-resolution telescopes like ACT or the Simons Observatory Large Aperture Telescope, but still performs reasonably well for the SO SAT.

\begin{figure}
    \centering
    \hspace{0.5cm}
    \includegraphics[width=\linewidth]{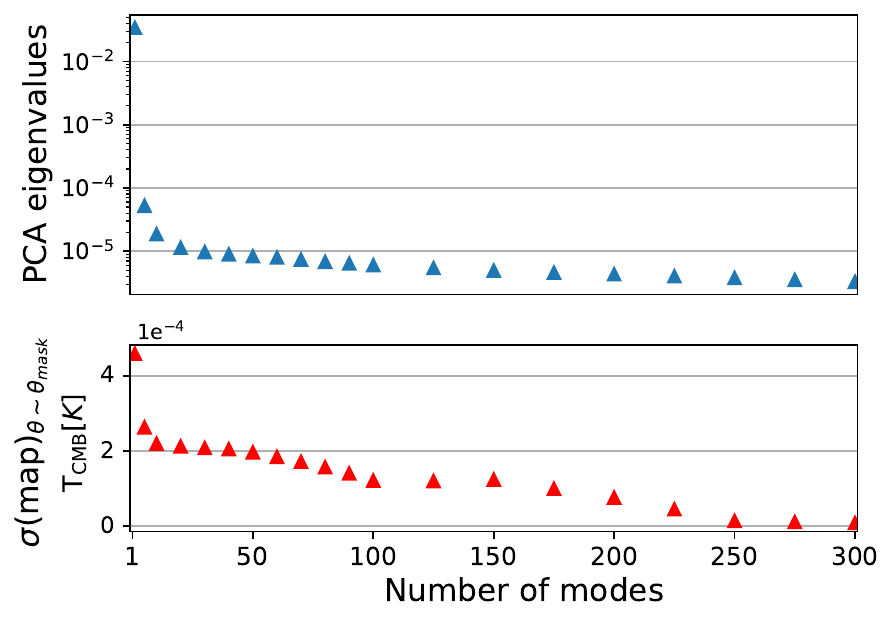} \\
    \caption{\label{fig:pca_weights} Top: The PCA eigenvalues of some of the strongest 300 correlated modes of the covariance matrix averaged over all detectors of the centre wafer for a single \SI{93}{\giga\hertz} Jupiter observation, as calculated from the map maker described in Section \ref{sec:map_making}. Bottom: The noise amplitude of the binned data as function of number of correlated modes subtracted. The noise level is calculated as the standard deviation of all the data at the outer 10$\%$ of the mask. Notice that the eigenvalues are shown in logarithmic scale while the noise levels are in linear scale.}
\end{figure}

\begin{figure*}[t!]
    \centering
    \includegraphics[width=0.8\textwidth]{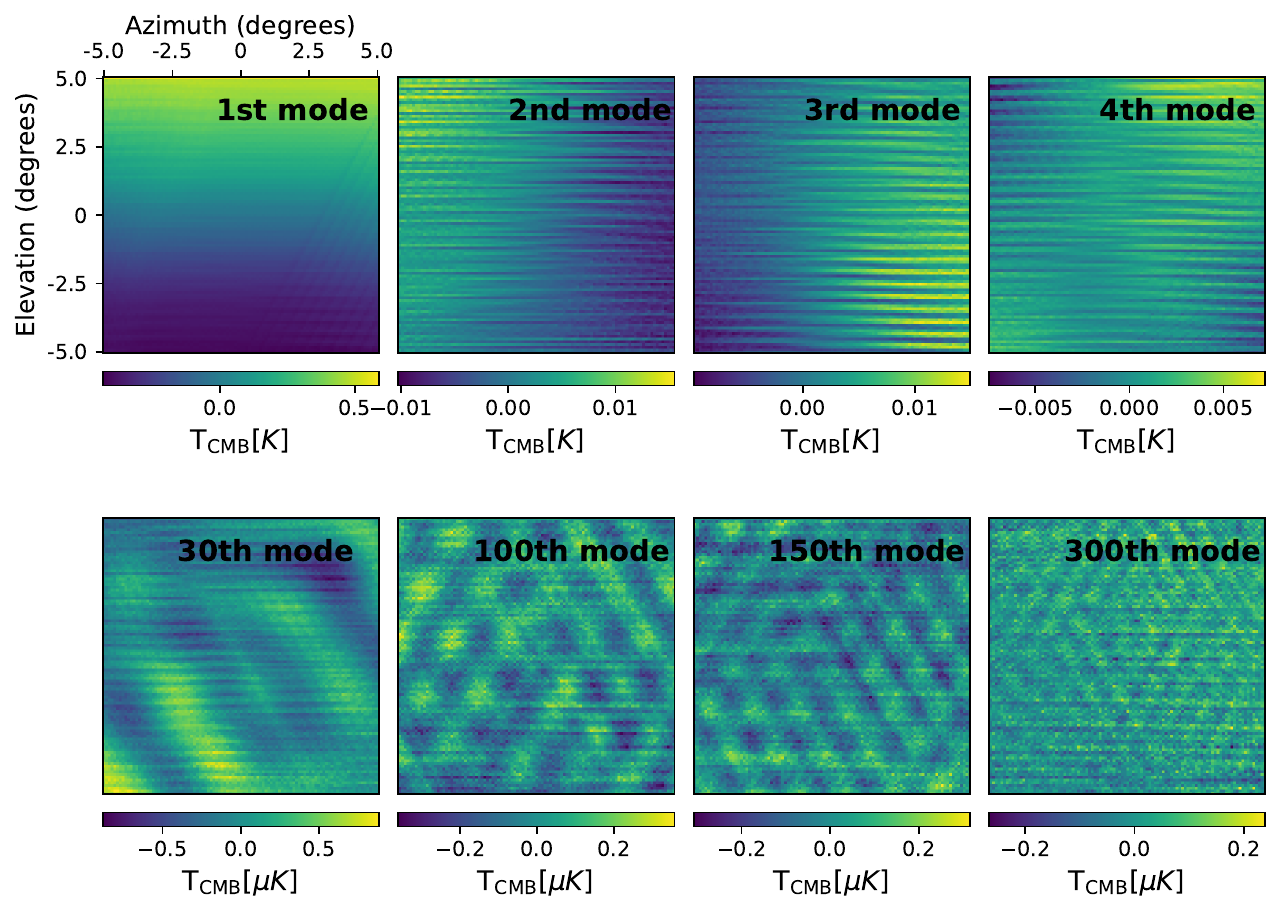} \\
    \caption{\label{fig:pca_modes} The binned \nth{1}, \nth{2}, \nth{3}, \nth{4}, \nth{30}, \nth{100}, \nth{150} and \nth{300} strongest correlated modes of the covariance matrix of the centre wafer detectors for a single \SI{93}{\giga\hertz} observation, as calculated from the map maker described in Section \ref{sec:map_making}. Please note the change of units from $K$ to $\mu K$ between the plots of the top and bottom rows.}
\end{figure*}

\begin{figure*}[t!]
    \centering    \includegraphics[width=0.82\textwidth]{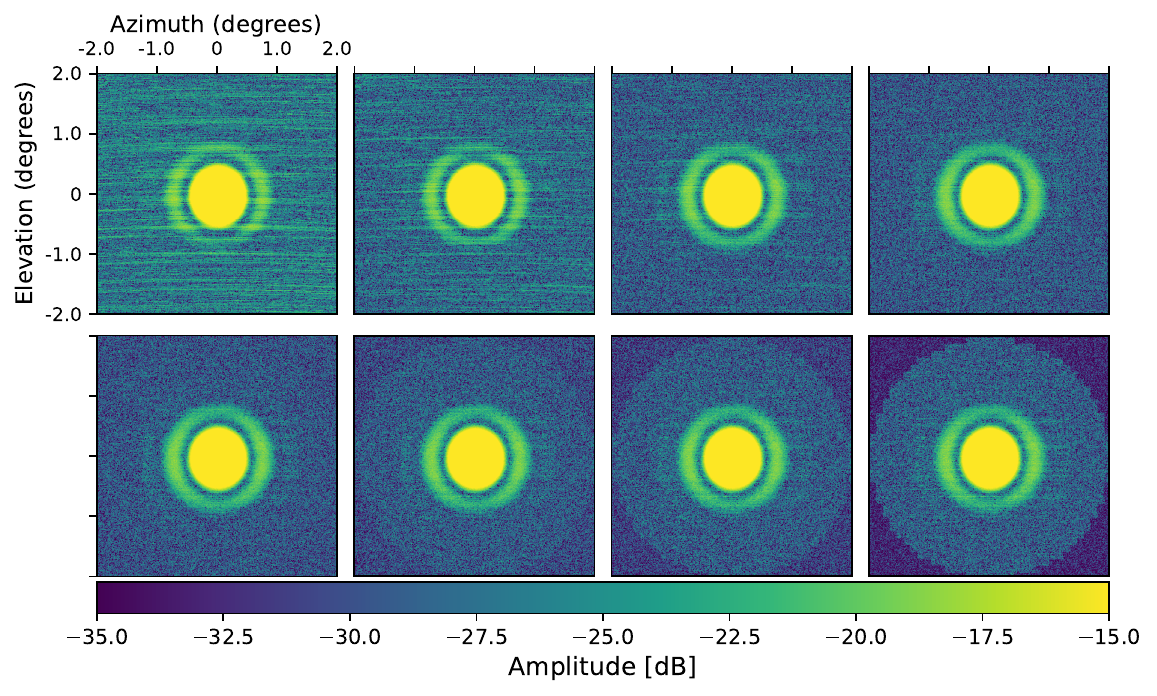} \\
    \caption{\label{fig:pca_maps} Peak-normalised beam maps constructed from the same \SI{93}{\giga\hertz} Jupiter simulation after subtracting up to the \nth{1}, \nth{2}, \nth{3}, \nth{4}, \nth{30}, \nth{100}, \nth{150} and \nth{300} strongest correlated modes shown in Figure \ref{fig:pca_modes}. Note that now we only show $\SI{4}{\degree} \times \SI{4}{\degree}$ patches around the source as we want to take a closer look at the noise subtraction effect inside the masked region. The chosen color scale saturates the beam but better captures the noise mitigation progression.}
\end{figure*}

\par The mask radius, $\theta_{\mathrm{mask}}$, around the source and the number of modes, $N_{\mathrm{modes}}$, can be tuned to optimize the performance of the algorithm. As the noise levels are estimated from the region that remains unmasked, a too small $\theta_{\mathrm{mask}}$ might result in subtracting beam power of substantial amplitude while one might fail to properly capture the relevant noise modes with a too large $\theta_{\mathrm{mask}}$. The top panel of Figure \ref{fig:pca_weights} shows the PCA eigenvalues of some of the 300 strongest modes for a single observation, performed with the \SI{93}{\giga\hertz} frequency band beam model and fixed PWV. Even though the total number of estimated modes equals the number of simulated detectors (860 detectors per single-wafer simulation), we find this truncated sample representative enough to capture the rate of the eigenvalues' decreasing amplitude. Each data point in this plot corresponds to the average eigenvalues of all the detectors of the centre wafer of the focal plane. The large value of the first mode presented in this figure ($\sim$2 orders of magnitude larger than the next mode) indicates how the atmospheric noise may be crudely approximated as a single correlated mode. The bottom panel of Figure \ref{fig:pca_weights} presents the noise amplitude of the same \SI{93}{\giga\hertz} planet map estimated as the standard deviation of all the data points included in the outer $10\%$ of the mask as a function of the number of subtracted modes. The results illustrate an overall reduction of the noise amplitude with increasing number of modes in an almost monotonic fashion. The noise variance is shown to be statistically compatible with zero after subtracting $\sim$ 300 modes.

The outer scale of atmospheric turbulence is observed to be significantly smaller than the 12-degree field of view of a single SAT wafer \citep{Errard_2015}. Splitting the wafer into subsets of fewer detectors and estimating the correlated modes across these subsets, instead, would likely adequately capture the atmospheric noise model with a smaller number of modes than in the case of estimating the noise correlations from the full wafer.
Relevant modifications to the map-making pipeline will be made, if necessary, when we start observations.

We set $\theta_\mathrm{mask}$ to be equal to a radius, outside of which the beam power has fallen below $\sim$ 0.01$\%$ of its peak value, following the example of \cite{Lungu_2022}. Figure \ref{fig:pca_modes} shows some of the binned modes that were calculated from the PCA analysis for the same simulation. The \nth{1} mode shows the atmospheric emission amplitude scaling with telescope boresight elevation, as expected, while the rest of the modes of the top row probe stripy patterns in slightly different directions and scales. The bottom row shows modes corresponding to detector correlations of significantly smaller amplitude. Subtracting these faintest modes might be excessive and could, in turn, end up negatively impacting the performance of the beam model reconstruction algorithm. Figure \ref{fig:pca_weights} suggests that the most suitable number of modes to subtract should be of the order of ten.

It is interesting to look at the impact of subtracting a different number of correlated modes directly on planet maps. Figure \ref{fig:pca_maps} shows a single, \SI{93}{\giga\hertz}, $\SI{4}{\degree}\times\SI{4}{\degree}$ planet map after subtracting the \nth{1}, \nth{2}, \nth{3}, \nth{4}, \nth{30}, \nth{100}, \nth{150} and \nth{300} mode, following the reasoning of Figure \ref{fig:pca_modes}. From this multi-panel plot, we see the atmospheric striping starting to subside after the $\sim$4th mode, while the contrast between the masked and unmasked region becomes more pronounced with increasing number of subtracted modes. As expected, the faintest eigenmodes of the covariance matrix will better capture the noise of the unmasked region since the eigenmodes are estimated in this region. Consequently, subtracting these faint modes from the maps will lower the noise amplitude in the unmasked region, but not impact as much the masked region around the source. 

\subsection{Beam profile fitting}
\label{sec:beam_profile_fitting}

For the beam profile fitting of the planet observations, we closely follow the method developed for the Atacama Cosmology Telescope (ACT)~\citep{Lungu_2022}. The beam fitting method is implemented in \texttt{beamlib}, a version of the code from the work of \cite{Lungu_2022} that has been adjusted to SO hardware specifications. This section summarizes the main aspects of the method.

The fitting pipeline takes a set of input maps and trims them to a size that should refer to a region well contained inside $\theta_{\mathrm{mask}}$. The next step is to correct for the bias caused by the noise mode subtraction. As in the case of the ACT beams, we find this effect to be fairly consistent with a constant offset deviation of the beam wing from the 1$/\theta^{3}$ function that the input SAT beams approximately follow (see Section \ref{sec:instrument}). For the constant offset estimation, we use a relatively `flat' part of the beam profiles (where the oscillatory sidelobe pattern is not as pronounced), and the beam power has fallen below $\sim 0.1\%$ of its peak value. The best-fit value for this offset is computed for each observation and then subtracted from each observation before averaging the beam profiles.

The core (main) beam is fitted, employing the basis functions from \cite{Lungu_2022}:
\begin{equation}
f_{n}(\theta \ell_{\mathrm{max}}) = \frac{J_{2n+1}(\theta \ell_{\mathrm{max}})}{\theta \ell_{\mathrm{max}}},
\end{equation}
where $J_{2n+1}$ are Bessel functions of the first kind, $n$ is a non-negative integer and $\ell_{\mathrm{max}}$ is allowed to vary around some mean value defined by the beam resolution. The angle of the transition from the `core' to the `wing' region of the beam is specified by the user, yet it is allowed to vary slightly in the code in order to retrieve its optimal value. The reconstructed beam transfer function is calculated as the Legendre transform of the best-fit model:

\begin{equation}
b_{\ell} = \frac{2\pi}{\Omega_{B}}\int_{-1}^{-1} B(\theta)P_{\ell}(\cos\theta)d(\cos\theta),
\end{equation}
where $P_{\ell}(\cos\theta)$ are the Legendre polynomials, $\Omega_{B}$ is the beam solid angle and $B(\theta)$ is the best-fit radial beam profile comprised of the core and wing fit.
Besides the best-fit harmonic transform, \texttt{beamlib} also calculates the eigenmodes of the beam-fitting covariance matrix, which we will refer to as error modes throughout the paper. 

\begin{figure}[t!]
    \centering
    \includegraphics[width=\linewidth]{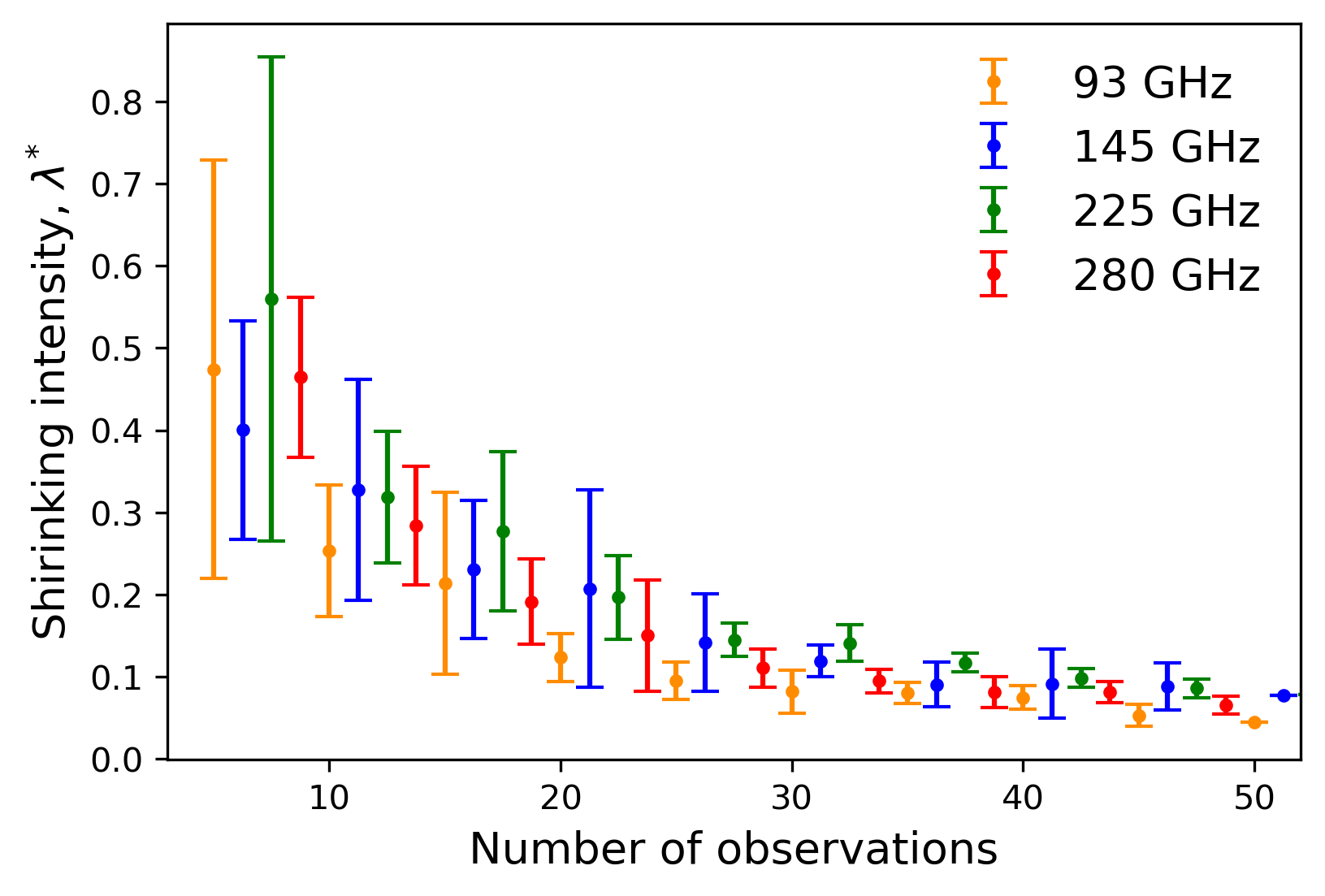}
    \caption{\label{fig:lambda_estimator} Shrinking estimator $\lambda^{*}$ as defined in \cite{Lungu_2022} as a function of the number of input Jupiter simulations for all frequency bands.}
\end{figure}

A small number of available planet observations can be problematic when estimating the bin-bin covariance matrix of the binned radial profile. For that reason, the code employs a shrinking technique. The idea behind the approach is the following: the level of down-weighting of the off-diagonal components of the covariance matrix should depend on the number of available observations. This approximation is parametrized by the so-called shrinkage intensity, $\hat{\lambda}^{*}$ (see Eq. \textcolor{blue}{(A5)}, Appendix \textcolor{blue}{A}, \cite{Lungu_2022}). The shrinkage intensity is applied to a biased version of the covariance matrix, \textbf{T}, where we have set all the off-diagonal components to zero and the empirical, unbiased covariance matrix, \textbf{S}, to synthesize the `shrunk' covariance matrix, \textbf{C}, that we will use for the beam fitting as (Eq. \textcolor{blue}{(A6)}, Appendix \textcolor{blue}{A}, \cite{Lungu_2022}):

\begin{equation}
\mathbf{C} =  \hat{\lambda}^{*}\mathbf{T} + (1-\hat{\lambda}^{*})\mathbf{S} 
\end{equation}

One can easily conclude that the larger the number of observations, the closer we can get to an unbiased covariance estimation. Figure \ref{fig:lambda_estimator} shows the value of $\lambda^{*}$ as a function of the number of input observations to the beam fitting code for our frequency bands. The shrinkage intensity seems to converge to a value of $\approx 0.1$ for a set of $N_{obs}$ = 50 observations in all cases.

\section{Results}
\label{sec:results}
We now present the reconstructed beam profiles and corresponding transfer functions for several different simulation parameters for the \SI{93}{\giga\hertz} band. A subset of indicative results is shown for the rest of the frequency bands as well. We are particularly interested in isolating the factors that will most strongly impact the quality of our beam reconstruction. 

\subsection{Dependence on the subtracted correlated noise modes}

The beam fitting performance is tightly linked to our ability to suppress the atmospheric signal sufficiently, which, in turn, depends on the number of correlated noise modes one removes from the data. For the simulation setup we employed, we find that the number of correlated modes that need to be removed typically lies between $N_{\mathrm{modes}}=5$ and $N_{\mathrm{modes}}=50$ for the frequency bands we consider.

\begin{figure}
    \centering
    \includegraphics[width=\linewidth]{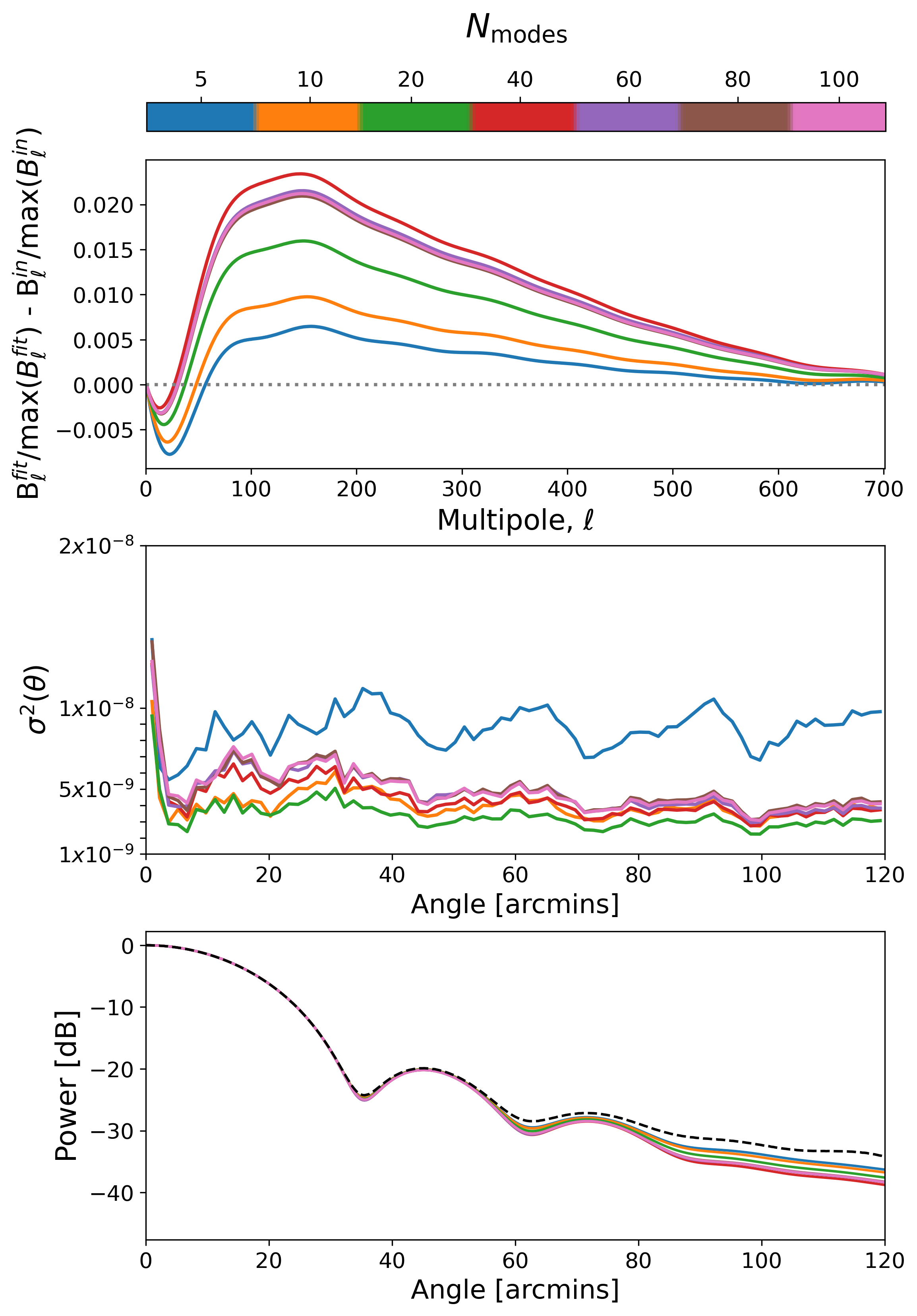} \\
    \caption{\label{fig:bias_var_fmodes} Top: The linear bias in multipole space between the normalized fitted and normalized input beam profile. Middle: the variance of the different planet maps (in logarithmic scale). Bottom: The best-fit beam profiles along with the input beam profile (black dashed line). All results are shown as a function of the number of correlated modes for the 93-GHz frequency band.}
\end{figure}

The choice of the number of subtracted modes relies on a combination of different criteria, as summarized in Figure~\ref{fig:bias_var_fmodes}. The top panel shows the linear bias in multipole-space between normalized fitted and normalized input beam transfer function as a function of the number of modes removed (varying colors) for simulations performed at the \SI{93}{\giga\hertz} frequency band. The middle panel shows the corresponding real-space variance between planet maps for the different number of modes subtracted. Each curve shown in this panel is estimated by computing the per radial bin variance of the beam profiles of a set of maps that have the same number of correlated modes removed. The bottom panel of the figure refers to the best-fit beam profiles of the same planet sets along with the input beam model.

From the middle panel, we see that the logarithmic variance between different observations reduces consistently with an increasing number of subtracted modes, but after some mode ($N_{\mathrm{modes}} \gtrsim 20$), it starts increasing again. We expect the strongest noise modes to be quite similar between simulations assuming different dates, but fainter modes should reflect some degree of day-to-day atmospheric changes. Subtracting these fainter modes inevitably increases the variance between different maps to some extent. Such behavior can indicate that we should not remove any more modes to avoid the risk of also subtracting signal along with the noise.  

The chosen number of modes, $N_{\mathrm{modes}}$, should be the best combination of low bias and low variance, which is the case for $N_{\mathrm{modes}}$=10 --20. Ideally, we would like the bias of the fitted beam transfer function to be fairly constant across the different multipole bins. We decide to use $N_{\mathrm{modes}}=10$, for the \SI{93}{\giga\hertz} case.

Higher frequency bands require more modes to be removed since the atmospheric brightness scales with frequency. For 145, 225, and \SI{280}{\giga\hertz}, we find that the number of modes that best satisfies our selection criteria is $N_{\mathrm{modes}}$ = 10, 30 and 40, respectively. Depending on the different simulation parameters, one might find a preferred (narrow) range of modes instead of a single global value. An alternative approach would be to subtract all the correlated modes and estimate a transfer function for the bias caused by this process by running additional simulations. For this work, we aim to only mildly bias our data so that the nature of this bias is fairly predictable and, in turn, easily corrected. 

\subsection{Dependence on detector position}
\label{sec:detector_position}
The fidelity of the beam reconstruction process depends on the input beam models. The input models depend, in turn, on the position of the detector on the focal plane. To assess the impact of the detector position on the fitting performance, we use beam models for a pixel on the centre and the edge of the focal plane. In both cases, we bin the data into maps with the ten most correlated modes removed after masking and gap-filling a region that extends to a radius of $\theta_\mathrm{mask}$ = \SI{2.5}{\degree} around the source. The gap-filling is done with polynomial interpolation of the unmasked data over the masked region.  

\begin{figure}[t!]
    \hspace{0.2cm}
    \includegraphics[width=\linewidth]{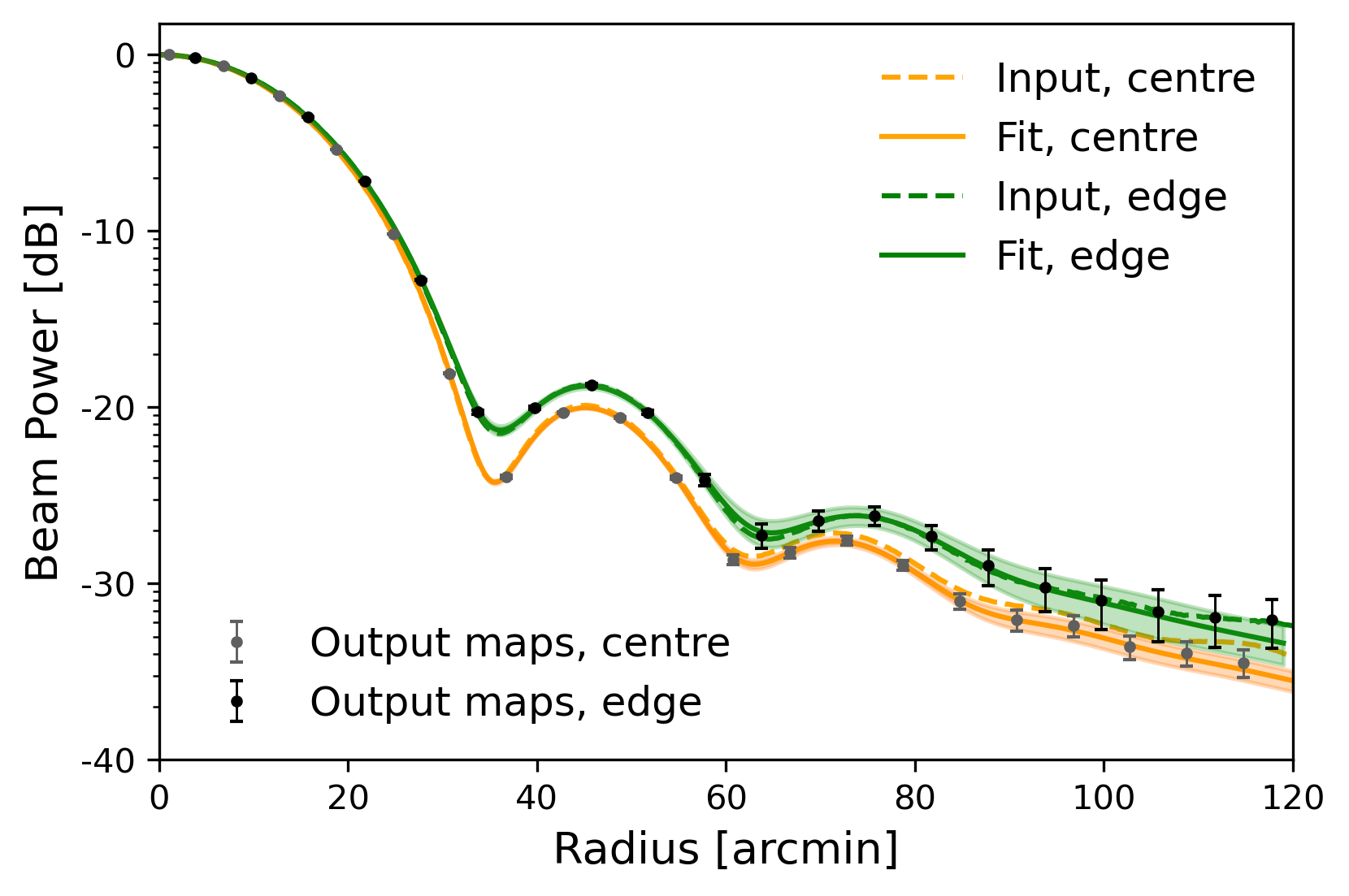} 
    \caption{ \label{fig:profiles_multi} Best-fit beam profiles generated from 3 months of Jupiter simulations for the \SI{93}{\giga\hertz} frequency band, performed with an input beam model for a centre (orange curve) and an edge pixel (green curve). The input beams (orange/green dashed lines) and data points (gray/black errorbars) of the fitted models are also shown.}

    \includegraphics[width=\linewidth]{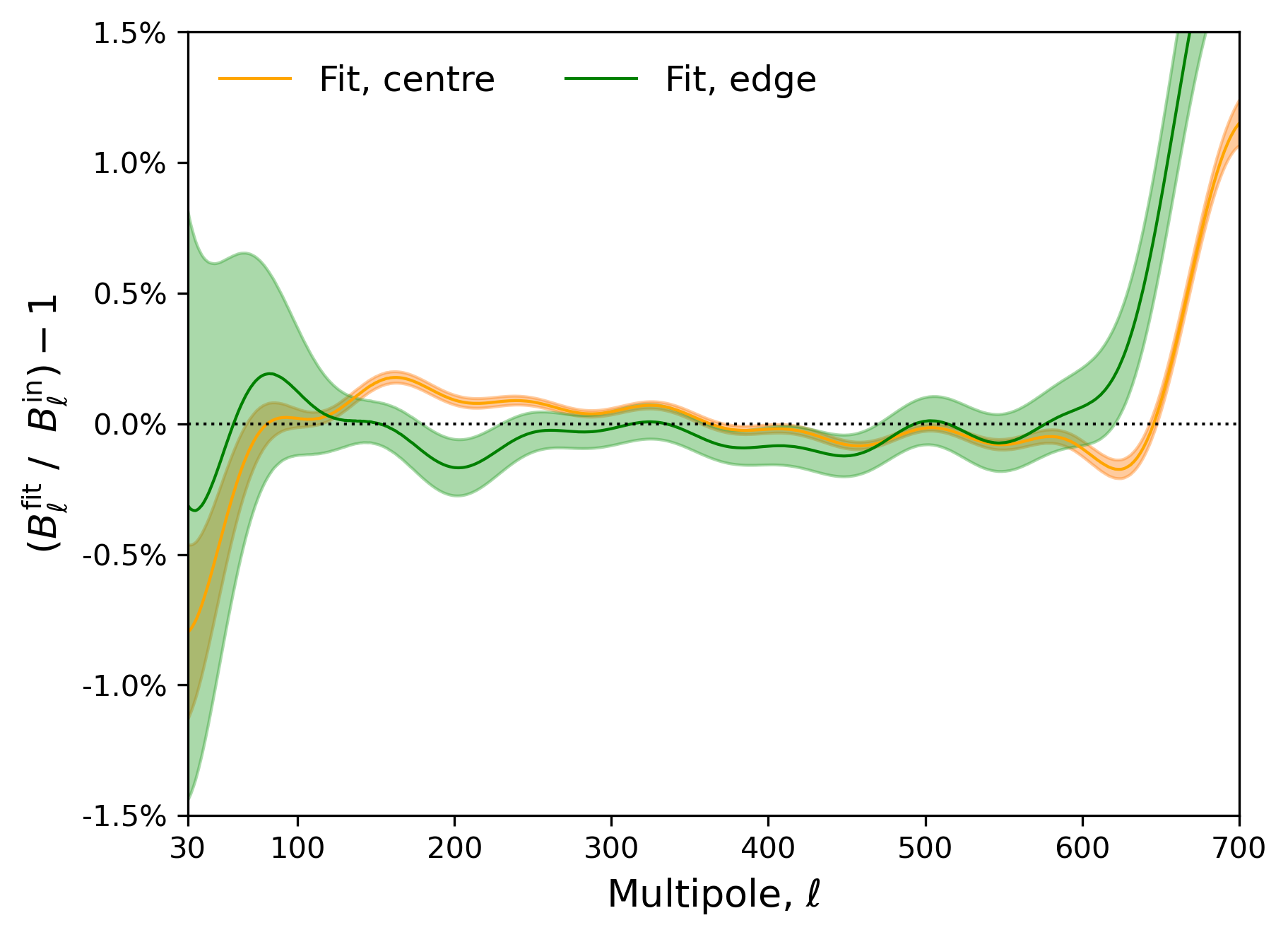}
    \caption{\label{fig:wf_multi} The beam transfer function bias, as compared to the input model, generated from 3 months of Jupiter simulations for the \SI{93}{\giga\hertz} frequency band, performed with an input beam model for a centre (orange curve) and an edge pixel (green curve). The band around the solid lines represents the $1\sigma$ error envelope determined from the beam error modes. 
    }
\end{figure}

Figure \ref{fig:profiles_multi} shows the reconstructed beam profiles from a set of three months of simulated Jupiter observations for the centre (orange curve) and one of the edge wafers (green curve). The detectors of the centre wafer share the centre pixel beam model while those of the edge wafer are assigned the edge pixel beam model (as discussed in Section \ref{sec:instrument}). The selection of the exact edge wafer is not important as the TICRA TOOLS setup is radially symmetric with respect to any of the edge pixels of the telescope's focal plane (see Figure \ref{fig:GRASP_setup}). The atmospheric PWV was set to $\sim$ \SI{1}{\milli\meter} at all times and we did not allow for wind speed variations. The results show the fitted centre-pixel beam model following closely the input model, up to $\sim$ 4 times the beam size (FWHM = \SI{27.4}{\arcmin} at \SI{93}{\giga\hertz}), while the edge-pixel beam profile deviates slightly from the input towards the largest radial bins. 

Figure \ref{fig:wf_multi} quantifies the above statement in terms of the bias on the beam transfer functions of the two reconstructed beam profiles compared to the input beams. From the figure, we can see that, while the centre-pixel model reconstruction is better, the transfer function bias remains well under $1.5 \%$ for a multipole range $\ell$ = 30 -- 700 for both cases. 

\subsection{Dependence on weather conditions}
\label{weather_conditions}

Different atmospheric conditions could affect the performance of the beam reconstruction algorithm. During the chosen observation period, we expect the weather conditions to vary to some extent. Generally speaking, there are a variety of parameters driving the atmospheric behavior. From these parameters, the amount of Precipitable Water Vapor (PWV) has the strongest impact on the atmospheric emission \citep[see e.g.,][]{D_nner_2012, Errard_2015}. 

The impact of PWV on atmospheric transmission can be seen in Figure \textcolor{blue}{1} of \cite{Errard_2015}. We define the transmission at some frequency, $\nu$, as the ratio $\mathcal{T}(\nu) \equiv I(\nu)/I_{0}(\nu)$ of the radiation received by the detector, $I(\nu)$, and the radiation above the atmosphere, $I_{0}(\nu)$. A high PWV value implies a low transmission $\mathcal{T}(\nu)$ which can be defined as the negative exponent of the airmass, $m(\SI{90}{\degree} - \mathrm{el})$, times some standard value of the optical depth, $\tau_{0}$, measured at the zenith \citep{Errard_2015}:

\begin{equation}
\mathcal{T}(\nu) = e^{-m(\SI{90}{\degree}-\mathrm{el})\tau_{0}},
\end{equation}
where $\mathrm{el}$ is the elevation. The airmass is, in turn, computed as a function of the zenith angle (see Equation \textcolor{blue}{(2)} of the same paper). This relationship holds at high elevations if we model the atmosphere as a parallel planar slab; in this case, $m(\SI{90}{\degree} - \mathrm{el}) \approx 1/\sin(\mathrm{el})$. The atmospheric transmission contributes to the total loading, $\mathcal{E}(\nu)$, as follows \citep{Errard_2015}: 

\begin{equation}
\mathcal{E}(\nu) = [1-\mathcal{T}(\nu)]B_{\nu}(T_{\mathrm{atm}}).    
\end{equation}
In the above, $B_{\nu}(T_{\mathrm{atm}})$ is the spectral radiance of a blackbody of temperature equal to the atmospheric temperature, $T_{\mathrm{atm}}$.  

\begin{figure}
    \centering
    \includegraphics[width=\linewidth]{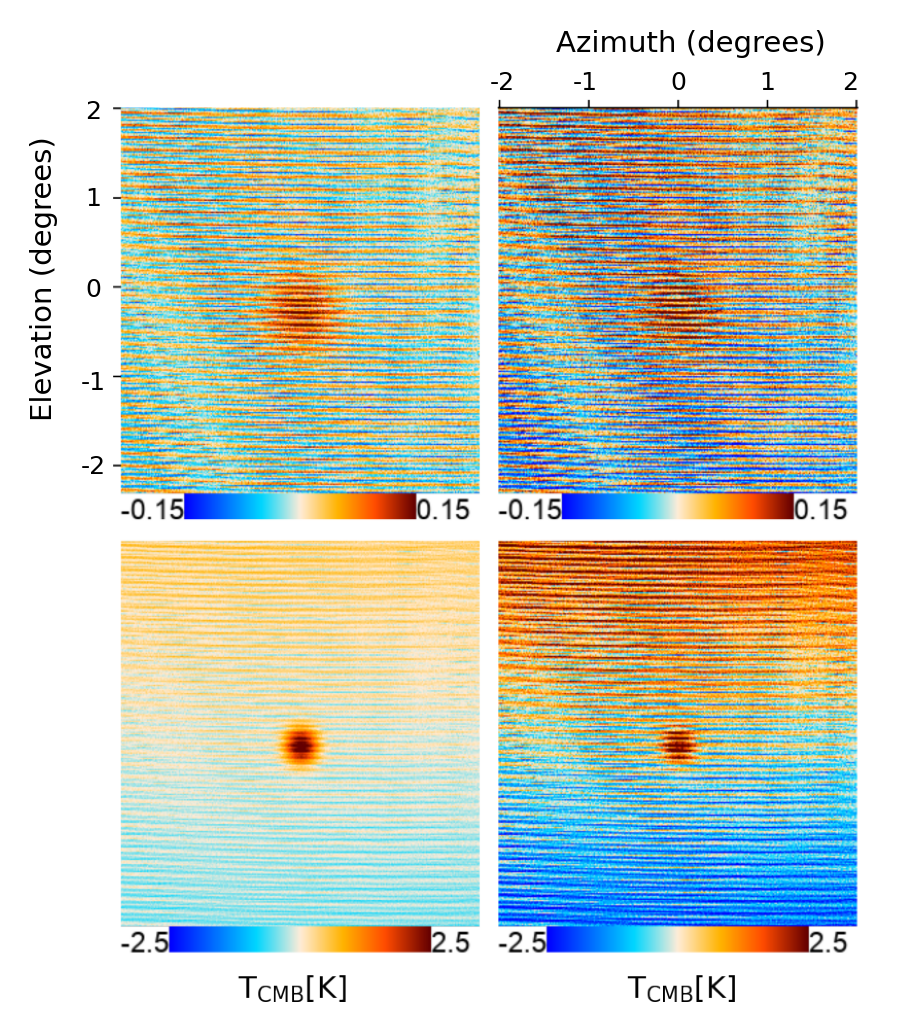}
    
    \caption{\label{fig:atm_pwv}Binned data of a single, $\sim$ 1-hour, Jupiter observation for all detectors in the centre wafer at 93 (top) and \SI{280}{\giga\hertz} (bottom) that have the mean temperature subtracted and have no correlated noise modes removed. The left and right panels represent simulations that include atmospheric emission of PWV = 0.5 and \SI{2.5}{\milli\meter}, respectively.}
\end{figure}

Figure \ref{fig:atm_pwv} shows the binned time-ordered data of a single Jupiter observation, including atmospheric emission of $\mathrm{PWV} =\SI{0.5}{\milli\meter}$ (left column) and $\mathrm{PWV} =\SI{2.5} {\milli\meter}$ (right column) at 93 (top row) and \SI{280}{\giga\hertz} (bottom row). The data are binned after subtracting the mean (atmospheric) temperature and have no correlated modes removed. The atmospheric intensity and therefore striping is shown to increase non-negligibly with PWV value, as expected. For the cases presented, the SNR at 0.5 (2.5) \SI{}{\milli\meter} is estimated as SNR = 35 (30) for the \SI{93}{\giga\hertz} band and SNR = 140 (60) for the \SI{280}{\giga\hertz} band, respectively.

The atmospheric signal is strongly correlated between different detectors. The wind speed and direction impact the correlation length along with the outer scale of turbulence and scan strategy. For two detectors with beam centroids that lie parallel to the wind direction, we will observe maximum signal correlation (Equations \textcolor{blue}{23-26} of \cite{Morris_2022}). To explore these effects, we run simulations for a center pixel in the \SI{93}{\giga\hertz} frequency band and different cases of atmospheric parameters described in Table \ref{tab:atm_params}. 

\begin{center}
\begin{table}[h!]
\centering
\begin{tabular}{ c c c c } 
 \hline
 Case & PWV & South wind speed & West wind speed \\  
 & [mm] & [m/s] & [m/s] \\ 
 \hline
 i & 1.17 & -1.25 & 3.4 \\
 ii & 1.17 & -1.25 $\pm$ 1 & 3.4 $\pm$ 2.5 \\ 
 iii & 2.5 & -1.25 & 3.4 \\ 
 iv & 1.17 $\pm$  1 & -1.25 & 3.4 \\ 
 v & 1.17 $\pm$  1 & -1.25 $\pm$ 1 & 3.4 $\pm$ 2.5 \\ 
 \hline
\end{tabular}
\caption{\label{tab:atm_params} PWV, wind speed, and direction assumed in the various simulation cases described in Section~\ref{weather_conditions}.}
\end{table}
\end{center}

The PWV value, wind speed, and direction are either fixed or allowed to fluctuate around a mean value following some distribution that is consistent with historical distributions according to MERRA-2 \citep{merra2} for the Atacama observation site. These distributions are included in \texttt{TOAST} and are specified per hour of the day and month. The mean and standard deviation values quoted in Table \ref{tab:atm_params} are synthesized from the individual simulations of the full observing period we have chosen.

 Notice that the estimated fixed mean PWV value strongly agrees with the one motivated by seasonal data of the ACT telescope ($\sim$ \SI{1}{\milli\meter}), which is located at the SO observation site (see Figure \textcolor{blue}{4} of \cite{Morris_2022}). The simulated PWV is uniformly distributed, and the surface temperature and pressure are kept constant at \SI{270}{\kelvin} and \SI{530}{\hecto\pascal}, respectively. The wind speed values in Table \ref{tab:atm_params} have a positive or negative sign in order to also incorporate the wind direction.

\begin{figure}
\centering
    \includegraphics[width=\linewidth]{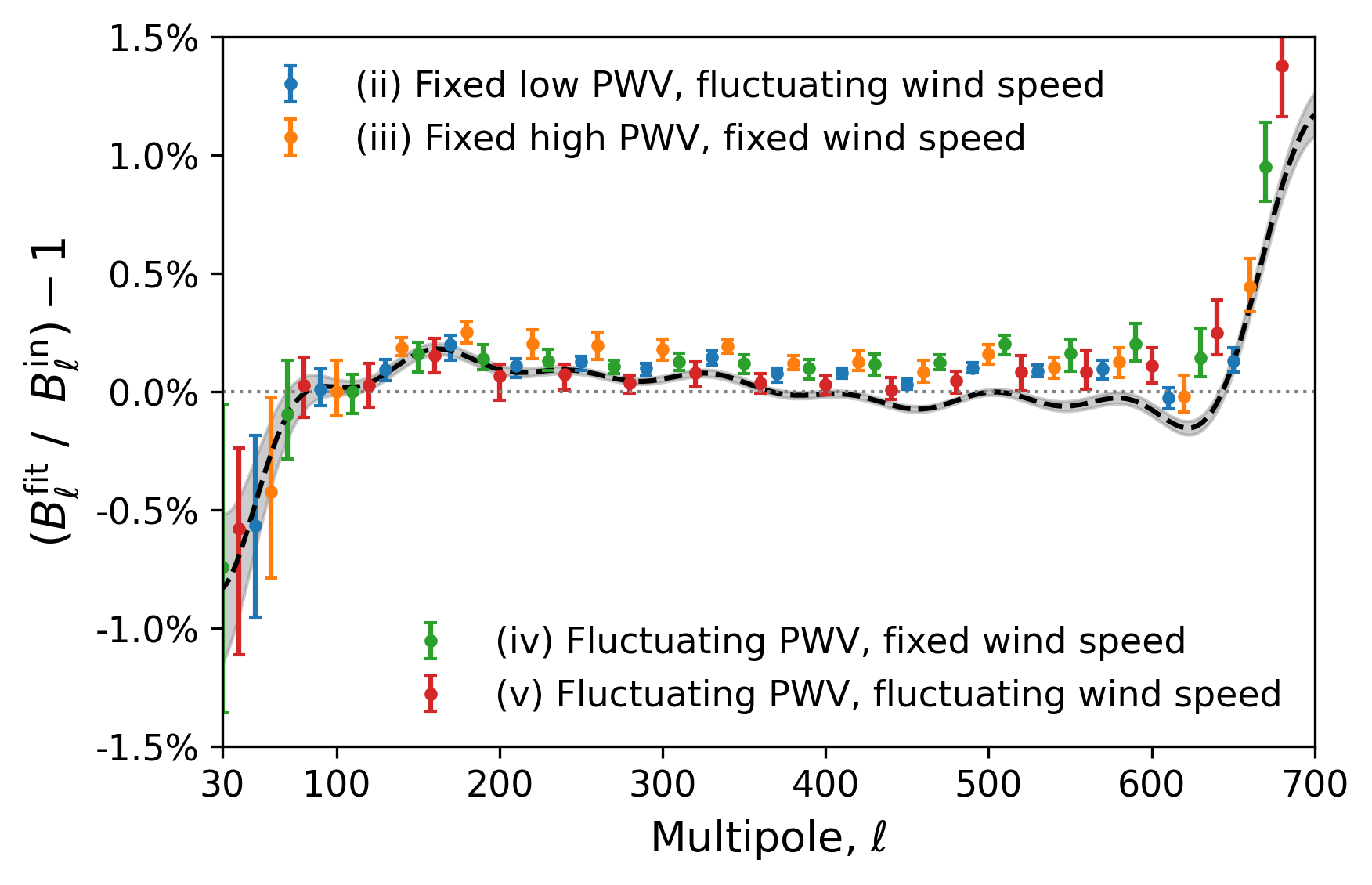} 
    \caption{ \label{fig:wf_weather} Beam transfer function bias from the input beam model, for the simulation cases (ii) - (v) described in Table \ref{tab:atm_params}. The simulations were performed with the \SI{93}{\giga\hertz} frequency band beam model assuming a pixel at the centre of the telescope's focal plane. Note that, for ease of visualization, the different errorbars are slightly offset in the $\ell$-direction. The black dashed case corresponds to the nominal case (i).}
\end{figure}

Figure ~\ref{fig:wf_weather} shows the uncertainty of the reconstructed beam transfer function with respect to the input beam model for the cases (ii), (iii), (iv) and (v), as described in Table~\ref{tab:atm_params}, in the form of blue, orange, green and red errorbars, respectively. The mean bias of the nominal case, (i), is demonstrated with a black dashed line surrounded by a gray-shaded uncertainty band for reference. The error bars are shown per 40 multipoles for easier visualization and extend to a multipole number of $\ell_{\mathrm{max}} = 700$. The beam fitting performance is rather stable across the different weather cases we have considered. However, the quality of the results slightly worsens with added atmospheric complexity, with the largest errorbars of Figure~\ref{fig:wf_weather} corresponding to the case where both PWV and wind speed are allowed to fluctuate (case (v)). A large PWV value implies an increased temperature of the atmospheric brightness. Since the latter also scales with frequency, the quality of the results depends on the ratio between the planet and atmospheric brightness at the frequency band of interest and how efficiently we can suppress the correlated noise. Nevertheless, we should highlight that, for all the atmospheric parameters chosen, we were still able to recover the input beam with an uncertainty smaller than $\sim 1.5 \%$ in all cases, for the multipole range $\ell$ = 30 -- 700 (for 93 GHz).

\subsection{Dependence on the number of observations}

The beam reconstruction algorithm depends on the number of available observations. To probe this, we present the accumulated SNR  as function of the number of Jupiter simulations for four different frequency bands centred on 93, 145, 225 and \SI{280}{\giga\hertz}. The beam models in the simulations assume a detector placed at the center of the focal plane. 

\begin{figure}[t!]
\includegraphics[width=\linewidth]{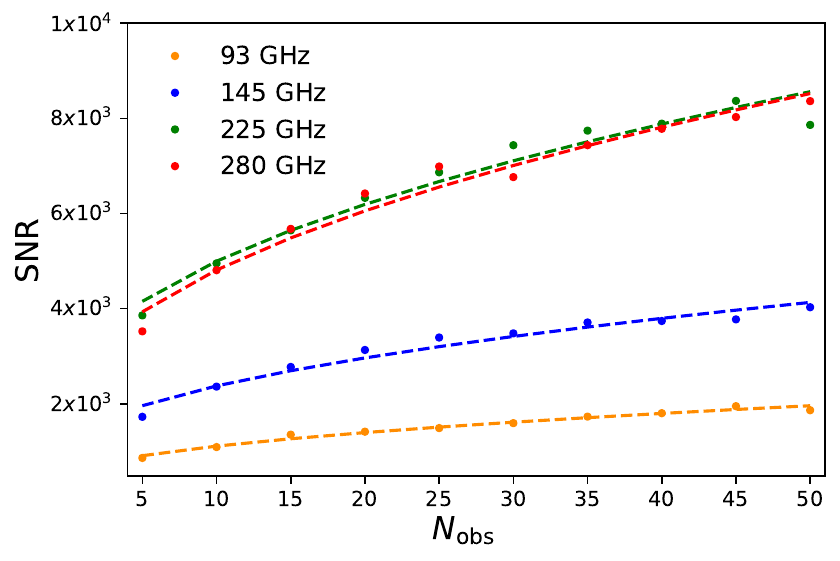}
\caption{\label{fig:all_freqs_snr}Signal-to-Noise ratio as a function of the number of simulated $\sim$ hour-long Jupiter observations for four frequency bands centred on 93, 145, 225 and \SI{280}{\giga\hertz}. The circles represent SNR values estimated by determining the noise levels of the maps, and the dashed lines represent the best fits of the data points to a $A\sqrt{N_{\mathrm{obs}}}+B$ model.}
\end{figure}

In our analysis, we face a trade-off between the overall accuracy of the reconstructed beam model and extending the model to larger angles. The SNR obviously decreases as we move away from the centre of the beam. Therefore, our attempts at fitting beam models in the faint wings of the sidelobes can sometimes bias our overall results. Assessing the reconstruction noise as a function of the number of observations is essential for optimizing the planet observing strategy. Figure \ref{fig:all_freqs_snr} shows the estimated SNR ratio for all frequency bands when the number of available observations ranges from 5 to 50. The dots refer to SNR values estimated by determining the noise that remains in the planet maps, which is approximated as the standard deviation of the data in the outer $10\%$ of the mask. The dashed lines are the fits of the SNR values to an underlying $A\sqrt{N_{\mathrm{obs}}}+B$ model (for some constants $A$ and $B$), which is the statistical behavior we would expect. Based on Figure \ref{fig:all_freqs_snr}, we decide that attempting to model all the frequency bands down to \SI{-35}{\decibel} is a reasonable choice. This value matches the acquired SNR for the lowest frequency band when the full observation set is employed and translates to a mask radius $\theta_{\mathrm{mask}} \sim$ 5 $\theta_{\mathrm{FWHM}}$ for the 93 GHz case. For consistency, we also use $\theta_{\mathrm{mask}} \sim$ 5 $\theta_{\mathrm{FWHM}}$ for the other frequency bands.

\subsection{Dependence on the frequency band}
\label{sec:all_freqs}

\begin{figure}[t!]
\centering
\includegraphics[width=0.91\linewidth]{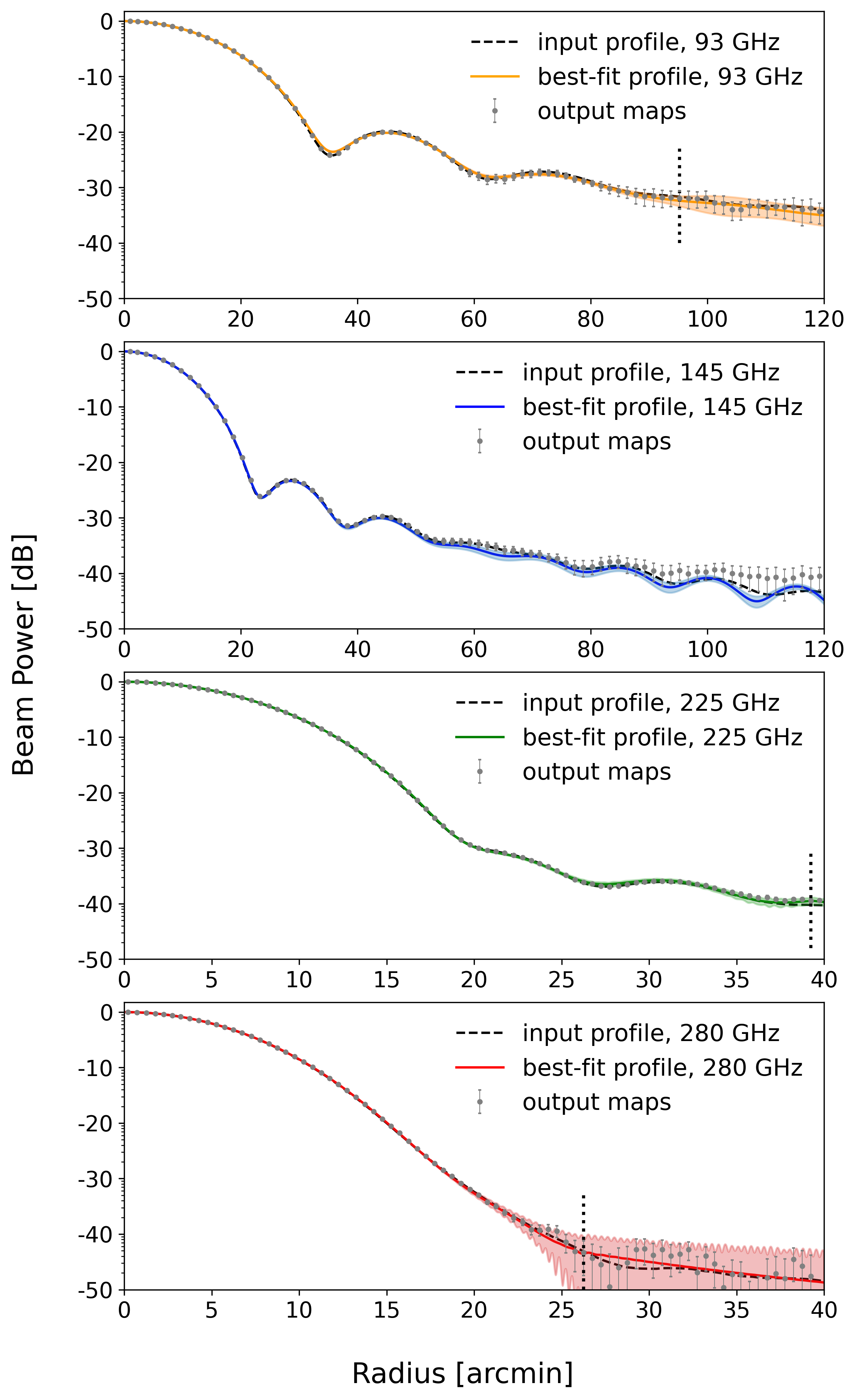}
\caption{\label{fig:all_freqs_profs} Best-fit beam profiles for the boresight pixel beam model at the 93-, 145-, 225- and \SIadj{280}{\giga\hertz} frequency bands. Note that the panels showing the 93- and \SIadj{145}{\giga\hertz} results have different horizontal range than those showing results for 225- and \SIadj{280}{\giga\hertz}. The vertical black line represents the transition between the core and wing fit.}
\centering
\includegraphics[width=0.91\linewidth]{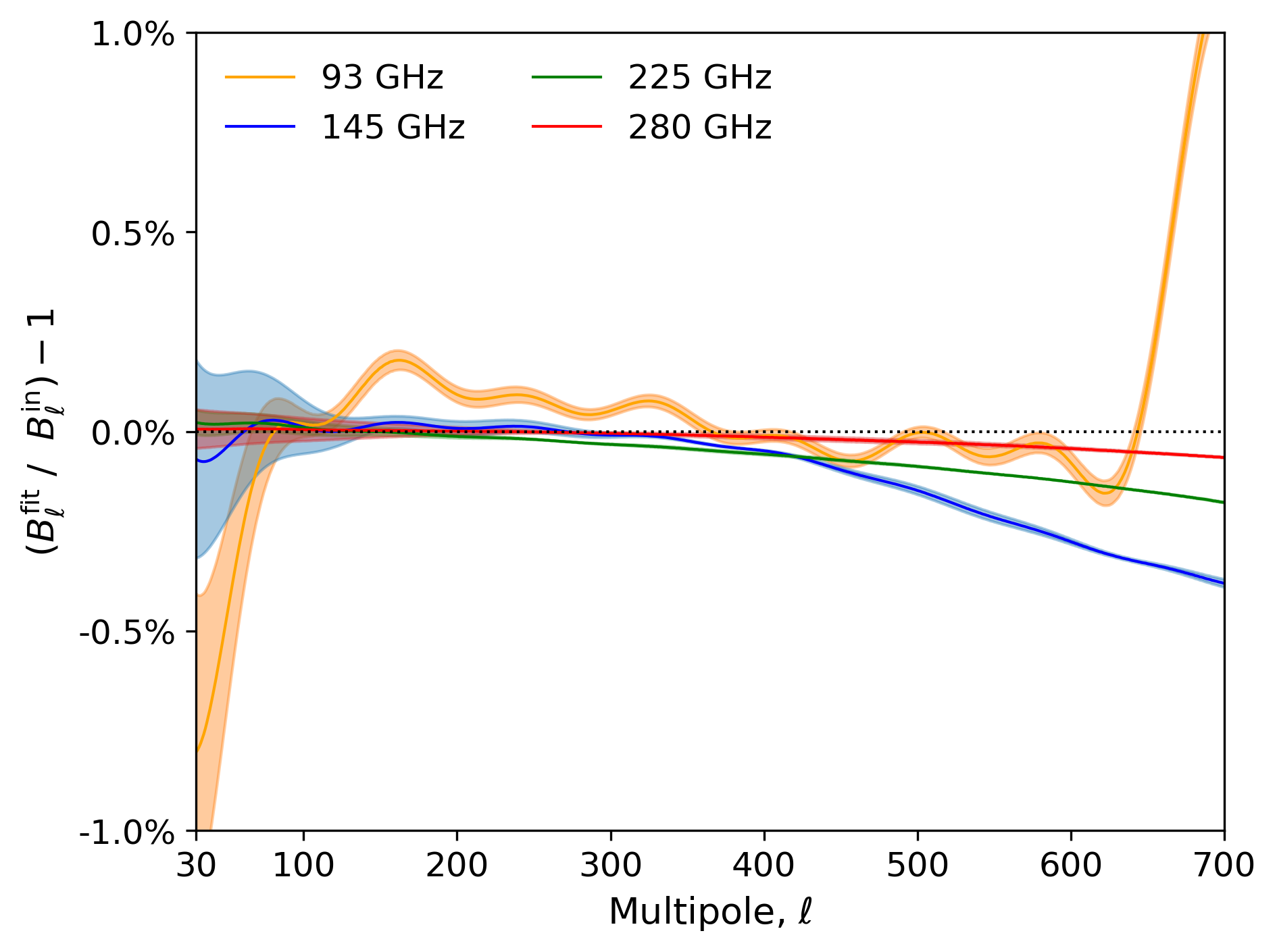} 
\caption{\label{fig:all_freqs_wfs_errors} Beam transfer function error with respect to the input beam per frequency band for the beam models of Figure \ref{fig:all_freqs_profs}.}
\end{figure}

Figure \ref{fig:all_freqs_profs} shows the reconstructed beam profiles for the four frequency bands compared to their input models. From the figure, we see the beam profiles of the MF and UHF bands following different sidelobe patterns in the target region. The \texttt{beamlib} code adapts to these differences by optimizing the interplay between the Bessel function basis model, which fits the beam core (where the sidelobe structure is expected to be more pronounced), and the $1/\theta ^3$ fit of the beam wing. The transition from core to wing fit is denoted with a black vertical line. Of all frequency bands, the 280-GHz band has the largest uncertainty on the reconstructed beam profile. The corresponding harmonic transform errors and their uncertainty, as calculated from \texttt{beamlib}, are presented in Figure \ref{fig:all_freqs_wfs_errors}. The plot shows a bias that roughly decreases in amplitude with increasing band centre frequency, especially for the multipole range 100 $< \ell <$ 300, and remains under $\sim 1.3 \%$ at all times. Table \ref{tab:btf_error} shows the maximum values of the reconstructed beam transfer function error, $\delta B_{\ell} = (B^{\mathrm{fit}}_{\ell}/B^{\mathrm{in}}_{\ell}) - 1$, for all frequency bands both for the full and a slightly truncated multipole range which will be further evaluated for calibration against Planck data in Section \ref{sec:cal_and_err_m}. 

The multipole region at which B-modes are expected to peak is still well contained within the truncated multipole range. The bias on the solid angle estimation, $\delta \Omega = (\Omega^{\mathrm{fit}}/\Omega^{\mathrm{in}})- 1$,  from \texttt{beamlib} is also shown.  

\begin{center}
\begin{table}[h!]
\centering
\begin{tabular}{ c c c c } 
 \hline
 Frequency & $\delta B_{\ell} $ & $\delta B_{\ell}$ & $\delta \Omega $ \\ band [GHz] & $\ell=30 - 700$ & $\ell=50 - 200$ &  \\ 
\hline
93  & $\le 1.2\%$ & $\le 0.6\%$ & $1.8\%$ \\ 
145 & $\le 0.4\%$ & $\le 0.2\%$ & $0.7\%$ \\ 
225 & $\le 0.2\%$ & $\le 0.06\%$ & $1.1\%$ \\ 
280 & $\le 0.1\%$ & $\le 0.05\%$ & $0.7\%$ \\ 
\hline
\end{tabular}
\caption{\label{tab:btf_error} The reconstructed beam transfer function and solid angle bias for the different frequency bands. The beam transfer function bias is shown both for the full and slightly truncated multipole range.}
\end{table}
\end{center}

These results reflect not only the expected scaling of the SNR as a function of frequency (and associated beam size) but also the success of the basis function choice for the beam model. This argument becomes evident when looking at the ringy pattern of the \SI{93}{\giga\hertz} band transfer function bias and associated uncertainty. Notice that the uncertainty of the reconstructed beam transfer function reduces as the number of available input simulations (and therefore accumulated SNR) increases. An estimate that quantifies this statement is provided in Appendix \ref{appendix:bl_f_nobs}.

\subsection{Calibration multipole range and error modes}
\label{sec:cal_and_err_m}
The technique chosen for the absolute calibration of the beam transfer functions will impact the $\ell$-dependence of the bias. Calibrating the SAT beam transfer function against previous CMB experiments,  such as Planck, is carried out by matching the spectra of the two telescopes over a limited range of multipoles. Since B-modes are expected to peak at a multipole number of $\ell \approx 80$, a calibration range around this lower multipole region is naturally motivated. 


We test the impact of different calibration choices directly on the bias of the reconstructed beam transfer function compared to the input. We do this by drawing $10^{4}$ realizations, $B'_{\ell}$, of the reconstructed beam transfer function $B_{\ell}^{fit}$ and the first 10 error modes $\delta B_{\ell}^{(i)}$:

\begin{equation}
\label{beam_realizations}
B'^{j}_{\ell} = B_{\ell}^{fit} + \sum_{i=1}^{10} c_{i,j}\delta B_{\ell}^{(i)}, \hspace{0.8cm} j=1, .., 10^{4}
\end{equation}
The weights, $c_{i,j}$, are randomly drawn from a normal distribution of zero mean and standard deviation equal to one. 

\begin{figure}
    \centering
    \includegraphics[width=\linewidth]{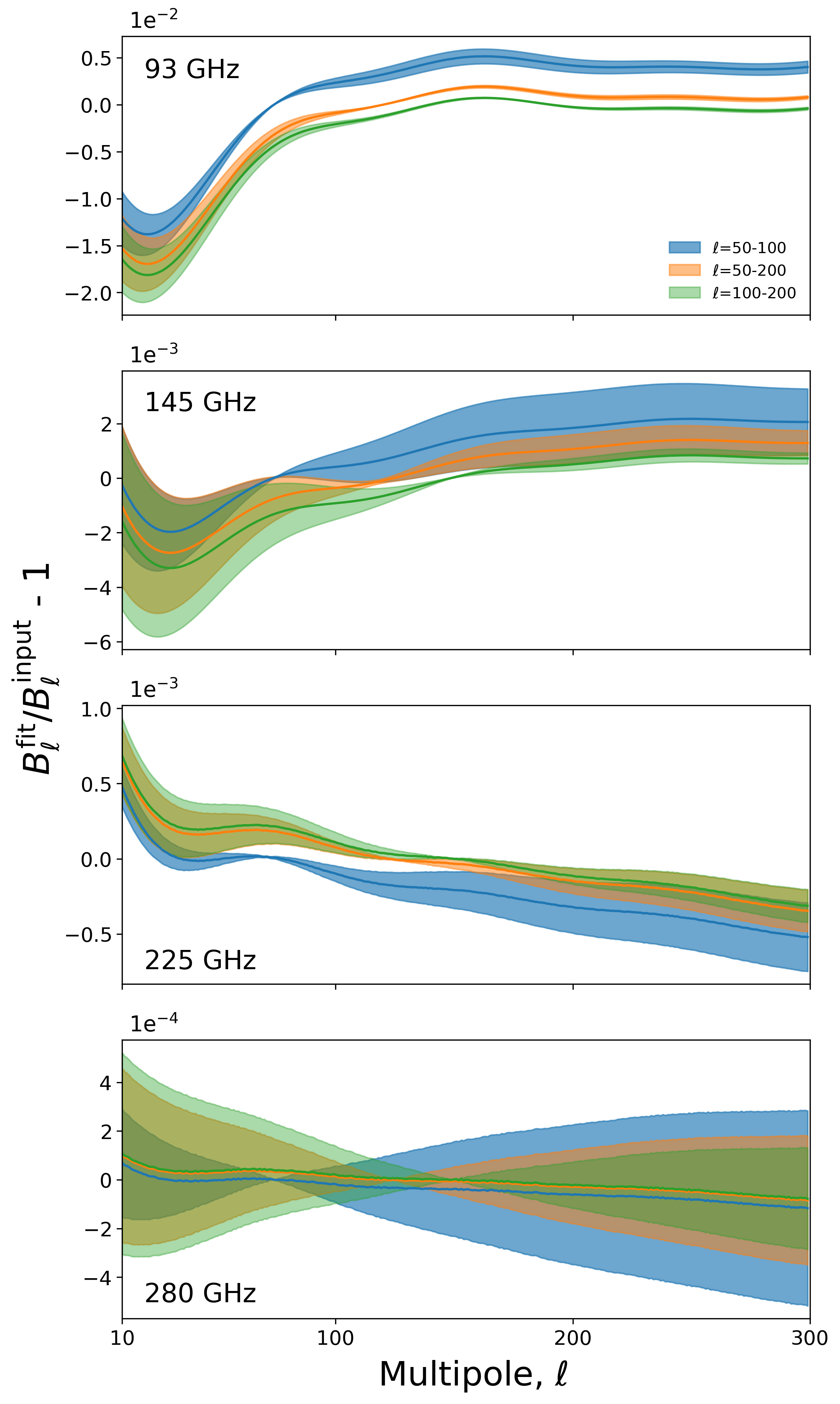} 
    \caption{\label{fig:cal_error} The 1$\sigma$-band of the beam transfer function bias with respect to the input beam for $10^{4}$ different beam realizations of the best-fit reconstructed beam and first 10 error modes for the four frequency bands centred on 93, 145, 225 and \SI{280}{\giga\hertz}. The beam realization transfer functions are calibrated on three different multipole ranges: $\ell$ = 50-100 (blue), $\ell$ = 50-200 (orange) and $\ell$ = 100-200 (green), respectively. All these ranges are suitable for calibrating the beam transfer function against experiments like Planck.}
\end{figure}

Given the SAT beams and expected transfer function, the multipole range that facilitates the absolute calibration of the SAT maps using the \textit{Planck} data will likely lie within $\ell_{\mathrm{min}}$ = 50 and $\ell_{\mathrm{max}}$ = 200. To investigate the impact of different choices of calibration range, we slice this multipole range and calibrate each one of the beam realizations we produced over the ranges $\ell$ = 50 -- 100, $\ell$ = 50 -- 200, and $\ell$ = 100 -- 200, by minimizing the difference between the output and input beam over each range. 
Figure \ref{fig:cal_error} shows the 1$\sigma$ error band of the $10^{4}$ newly calibrated beams, divided by the input beam transfer function, for the three multipole range choices quoted above. The results are shown for all four frequency bands (increasing in frequency from the top to the bottom plot) and are truncated to $\ell $ = 10 -- 300 for visualization purposes. As one can conclude from the plots, assuming a calibration range of $\ell$ = 50 -- 100 results in the minimum beam uncertainty at the low-multipole range of interest, while a calibration range at higher multipoles significantly increases the beam uncertainty at low multipoles.  

\subsection{Beam reconstruction uncertainty impact on the r-constraint}

The Simons Observatory Small Aperture Telescopes allow us to constrain the tensor-to-scalar ratio, $r$, with a statistical error of $\sigma(r) \le 0.003$. Beam modeling errors can bias cosmological analysis and it is therefore appropriate to briefly consider their impact on the forecasted value of $r$. For this purpose, we employ the \texttt{BBpower} software,\footnote{\textcolor{blue}{https://github.com/simonsobs/BBPower}} which is part of the publicly available SO analysis pipeline \citep{wolz2023simons}. \texttt{BBpower} is a harmonic-based component separation algorithm that has been adapted to the specifications of the SO telescopes \citep{Ade2019}. We use the code to forecast sensitivity to the value of the tensor-to-scalar ratio through Fisher analysis \citep{Fisher_1922saa}.

To quantify the effect of the beam reconstruction bias and uncertainty, we use Equation \ref{beam_realizations} to create 100 biased beam realizations for each of the four frequency bands considered in our analysis. For each beam realization, we construct a set of beam-convolved CMB and foreground spectra assuming no primordial B-modes ($r = 0$). The sky component and noise power spectra follow the ``Pipeline A'' with ``baseline'' noise level and ``optimistic'' $1/f$ noise description in \cite{wolz2023simons}. We then forecast the reconstructed $r$ value for each of the 100 realizations by (incorrectly) using the unbiased input beam to perform beam deconvolution in the fisher forecast code.
The resulting bias on the tensor-to-scalar ratio is $\Delta r = 1.08\cdot10^{-4}$. This number can be compared to the expected $1\sigma$ error on $r$, which is $\sigma (r) \approx 3\cdot10^{-3}$ \citep{Ade2019}. These beam errors add insignificantly to the overall variance on $r$: $\sigma(r)^{\mathrm{extra}} \sim 10^{-6}$. We thus conclude that the beam reconstruction error achieved with the setup presented in this work will be small enough to not significantly bias the SO $r$-measurement.
 

\section{Conclusions and Discussion}
\label{sec:conclusion}

This paper describes a beam reconstruction pipeline for the SO SAT beams in the MF and UHF frequency bands. The Low-Frequency (LF) bands are left for future work. We generate 50 $\sim$ one-hour-long Constant Elevation Scan (CES) simulations of Jupiter observations (as described in Section \ref{sec:scan_strategy}) and feed them to a filter-and-bin mapmaker designed to mitigate the correlated atmospheric noise by removing the strongest modes calculated from a Principal Component Analysis (PCA). The maps produced in this way are inputs to a slightly modified version of the ACT beam fitting code, \texttt{beamlib}. From this code, we obtain the best-fit beam profiles, transfer functions, and associated error modes. We present results that quantify the success of our beam fitting method as a function of different input beam models, weather, and frequency bands. These simulations allow us to assess the overall robustness of our analysis pipeline and prepare for the arrival of real data. 

Our simulations for the \SI{93}{\giga\hertz} band show that the beam reconstruction is generally robust to optical effects caused by detector location on the focal plane; we are able to reconstruct beam transfer functions with error not exceeding \SI{1.5}{\percent} in the $\ell = 30$--700 range and better than \SI{0.6}{\percent} in the $\ell = 50$--200 range. Testing how beam reconstruction for the \SI{93}{\giga\hertz} band depends on weather parameters shows similar results. This indicates that planet observations are useful even under relatively adverse weather conditions. 

The fitted beam profiles and transfer functions vary as a function of frequency. We model all four frequency bands to at least $\sim$ -35 dB and estimate the transfer function bias. The results show the fitting model adapting well to the different sidelobe patterns for the MF and UHF bands and the beam transfer function bias decreasing with increasing frequency. The uncertainty in the beam reconstruction can be reduced by optimizing the range of $\ell$ used to calibrate the data by comparing it to previous experiments (see Section \ref{sec:cal_and_err_m}). We find the preferred multipole range to be $\ell$ = 50--100 as it provides the lowest uncertainty on the beam transfer function over the $\ell $ = 10--300 region. 

We note that in the beam reconstruction error analysis marginalization over ad hoc choices, such as the wing scale and the number of subtracted modes, was not included and that this is different from what was done in \cite{Lungu_2022}. We expect these sources of error to somewhat increase the beam reconstruction uncertainty, particularly in the low-$\ell$ regime, but leave this analysis for future work.

Using simulated planet observations with a realistic atmospheric component, we observe beam reconstruction biases that are non-negligible compared to the uncertainty estimates (see Section \ref{sec:results}). However, these multiplicative biases are still relatively small ($<$ \SI{0.6}{\percent} in the $\ell = 50$--200 range for all the cases we tested) and are not expected to significantly impact the cosmological analysis. To verify this, we used a Fisher analysis to propagate the beam reconstruction bias and uncertainty. The result indicates that the reconstruction bias will be small enough to not significantly bias the SO $r$-constraint.

\section{Acknowledgments}

 ND and JEG acknowledge support from the Swedish National Space Agency (SNSA/Rymdstyrelsen) and the Swedish Research Council (Reg.\ no.\ 2019-03959). JEG also acknowledges support from the European Union (ERC, CMBeam, 101040169). The Flatiron Institute is supported by the Simons Foundation. GC is supported by the European Research Council under the Marie Sklodowska Curie actions through the Individual European Fellowship No. 892174 PROTOCALC. This manuscript has been authored by Fermi Research Alliance, LLC under Contract No. DE-AC02-07CH11359 with the U.S. Department of Energy, Office of Science, Office of High Energy Physics. The work was supported by a grant from the Simons Foundation (CCA 918271, PBL). DA acknowledges support from the Beecroft Trust, and from the Science and Technology Facilities Council through an Ernest Rutherford Fellowship, grant reference ST/P004474. SA is funded by a Kavli/IPMU doctoral studentship. GF acknowledges the support of the European Research Council under the Marie Sk\l{}odowska Curie actions through the Individual Global Fellowship No.$\sim$ 892401 PiCOGAMBAS. RG would like to acknowledge support from the University of Southern California. AHJ acknowledges support from STFC and UKRI in the UK. LP acknowledges support from the Wilkinson and Misrahi Funds. KW is funded by a SISSA PhD fellowship and acknowledges support from the COSMOS Network of the Italian Space Agency and the InDark Initiative of the National Institute for Nuclear Physics (INFN). ZX is supported by the Gordon and Betty Moore Foundation through grant GBMF5215 to the Massachusetts Institute of Technology.



\newpage
\appendix
\section{Varying input beam}
\subsection{Passband variations}
\label{appendix:var_in_beam}
\begin{figure}[h!]
    \centering
    \includegraphics[width=0.8\textwidth]{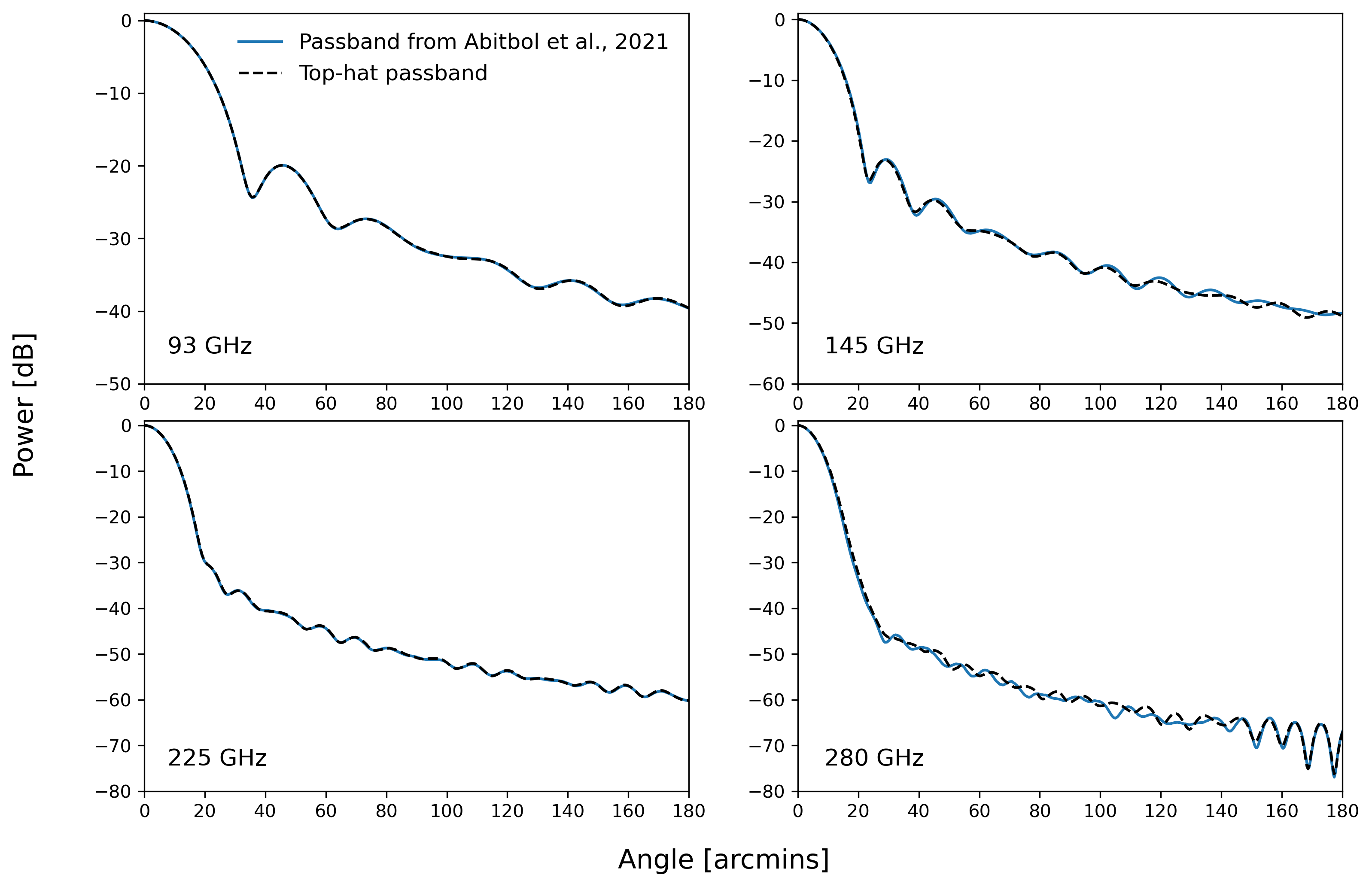}
    \caption{\label{fig:diff_bandpass} Logarithmic profiles of beam models where a top-hat (black dashed line) and a more realistic (blue line) passband was assumed. The realistic passband was taken from \cite{Abitbol_2021}, and the results refer to four frequency bands centred on 93, 145, 225, and \SI{280}{\giga\hertz}.}
\end{figure}

Variations in the detector passband will impact the shape of the effective beam and therefore affect the performance of the fitting algorithm. While we leave detailed analysis of this phenomenon for future work, it is useful to show how the input beam models may change under the assumption of non-uniform passbands. As stated in Section \ref{sec:instrument}, the frequency-band beams are produced by combining five monochromatic beams within a 20$\%$ bandwidth around the centre frequency with a top-hat passband. We compare the profile of the beams that were constructed this way to the ones where, instead of a top-hat function, we employed the simulated passbands from \cite{Abitbol_2021} and show the results in Figure \ref{fig:diff_bandpass} for all frequency bands.

Any difference between the uniformly and non-uniformly weighted beam profiles is negligible, at least to the $\sim \SI{-35}{\decibel} $ level we have chosen for fitting the SAT beams. Consequently, there is no indication from these plots that any change to the beam fitting method would be necessary. The passband assumptions/simulations we make will eventually be replaced with Fourier-Transform Spectroscopy (FTS) measurements to characterize the instrument’s spectral response. These measurements will enable us to produce realistic SAT beam models for future analysis.

\subsection{Beam chromaticity}
\label{appendix:var_in_sed}

\begin{figure}[h!]
    \centering
    \includegraphics[width=0.8\textwidth]{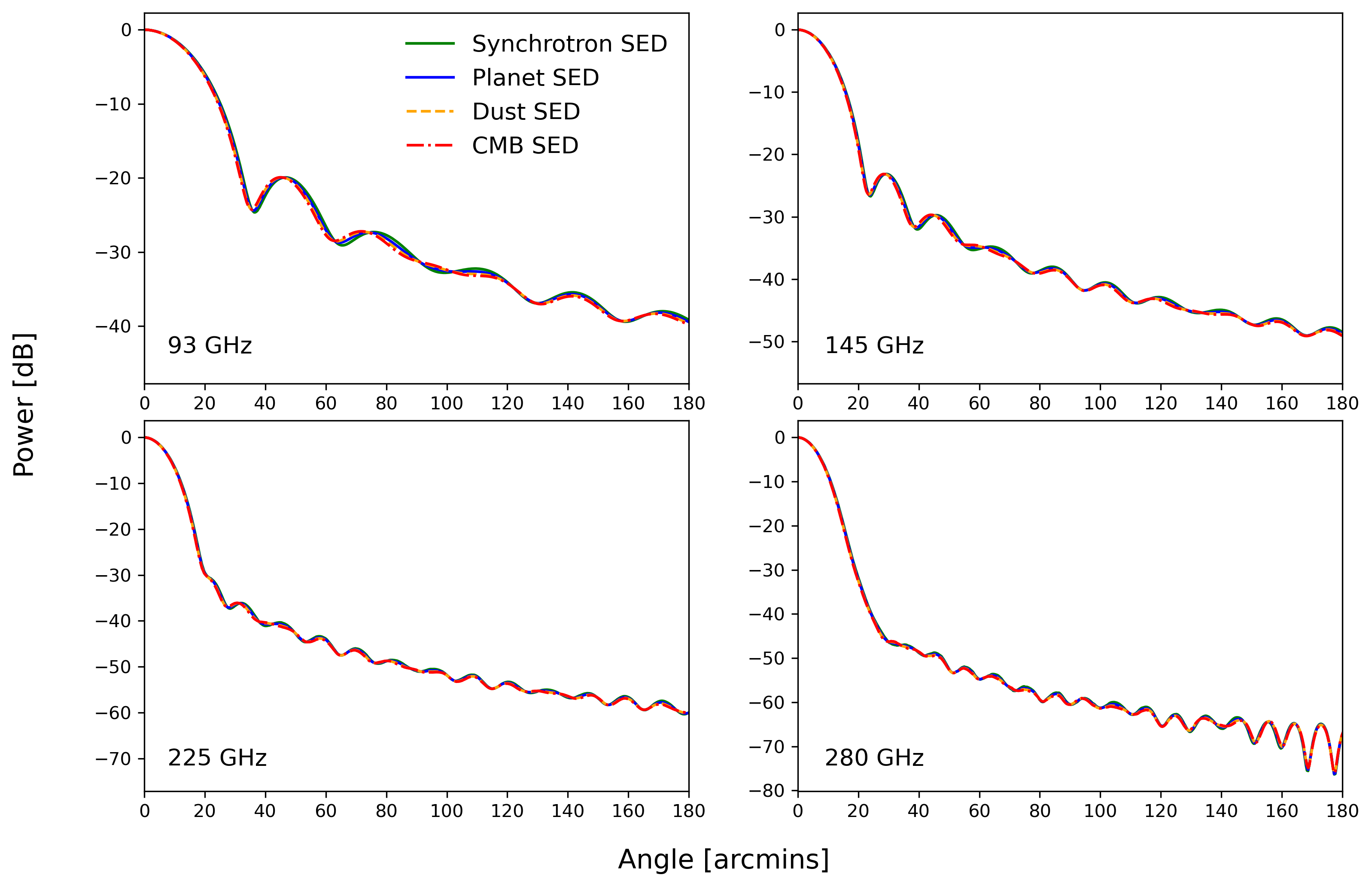}
    \caption{\label{fig:sed_profiles} Beam profiles for four models constructed using Equations \ref{beam_chrom1} and \ref{beam_chrom2} where the frequency scaling matches the SED of CMB (red curve), planets (blue curve), galactic dust (orange curve) and synchrotron emission (green curve). The plots refer to four frequency bands centred on 93, 145, 225, and \SI{280}{\giga\hertz} and include the case where no frequency scaling was implemented (black dashed line), for reference.}
\end{figure}

The instrumental beam can be frequency-scaled in a way that matches the SED of the different sky components that the telescope observes. Properly accounting for this effect is important for the performance of foreground component separation algorithms. Assuming a known passband, $\tau(\nu)$, of the instrument, the frequency-averaged beam, $B(\theta, \phi)$, can be described as:

\begin{equation}
\label{beam_chrom1}
B(\theta, \phi) = \int B(\theta, \phi, \nu) \tau(\nu) S(\nu) d\nu,  
\end{equation}
where $B(\theta, \phi, \nu)$ is a monochromatic beam at frequency $\nu$ and $S(\nu)$ captures the assumed frequency scaling. For many astrophysical sources, the latter can be expressed as a power law:
\begin{equation}
\label{beam_chrom2}
S(\nu) = \left(\frac{\nu}{\nu_{c}} \right) ^{\beta},
\end{equation}
where $\nu_{c}$ is the frequency band centre and $\beta$ is the spectral index. We consider four cases of frequency scaling matching the SED of CMB, planets, galactic dust, and synchrotron emission, corresponding to spectral index values of $\beta_{\mathrm{CMB}}$ = 1, $\beta_{\mathrm{planet}}$ = 2, $\beta_{\mathrm{dust}}$ = 1.56 and $\beta_{\mathrm{sync}}$ = -3 \citep{{planck_diff_sep_comp_2020}}. The beam profiles for these four cases, along with the case where no frequency scaling was implemented, are shown in Figure \ref{fig:sed_profiles} for all four frequency bands.  

\begin{figure}[h!]
    \centering
    \includegraphics[width=0.8\textwidth]{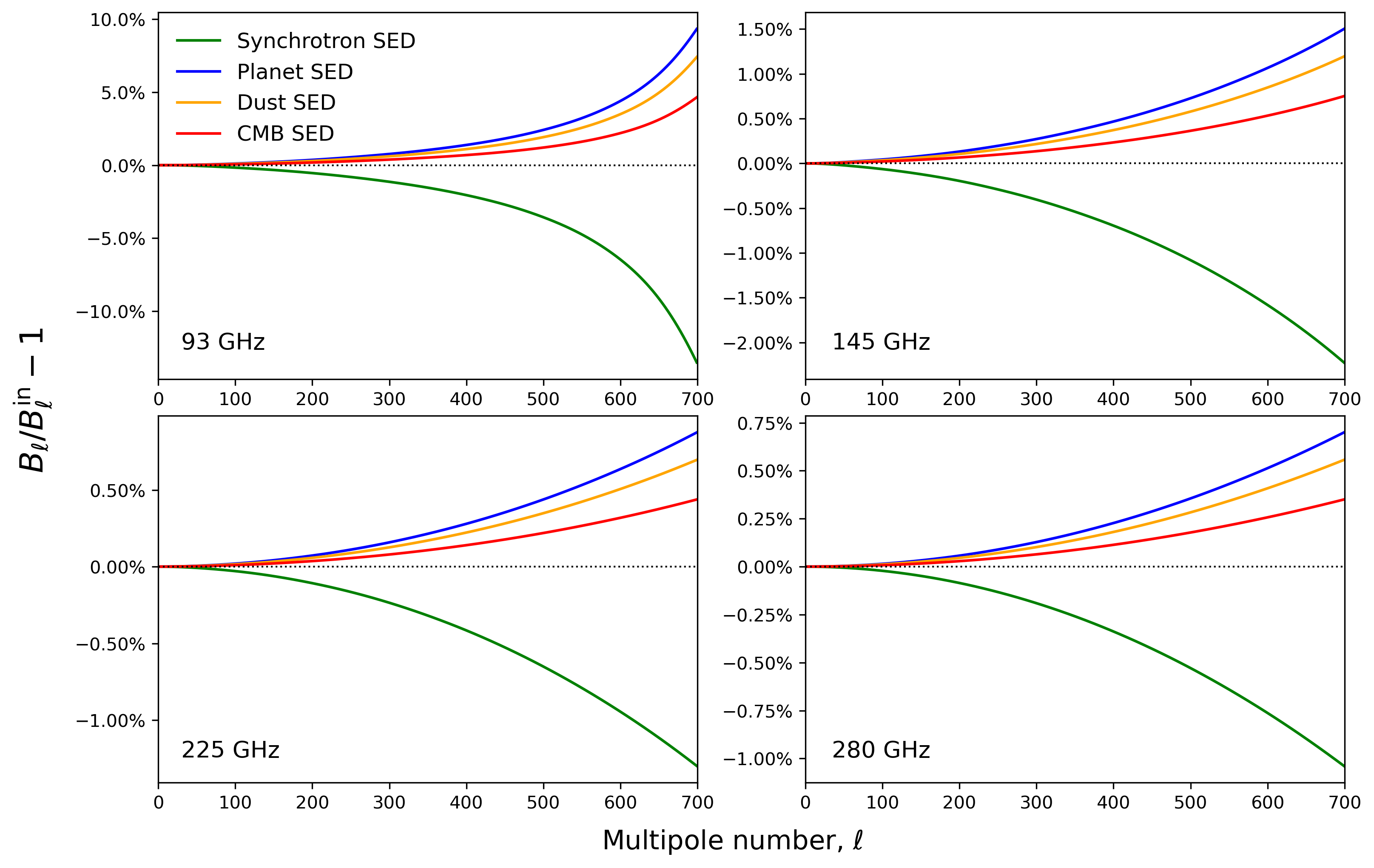}
    \caption{\label{fig:sed_wf_bias} The ratio of the beam transfer functions for the four chromatic beams whose profiles were shown in Figure \ref{fig:sed_profiles}, and the transfer function of the nominal case where no frequency scaling was implemented. The results are shown for four frequency bands centred on 93, 145, 225, and \SI{280}{\giga\hertz}.}
\end{figure}

It is interesting to see how the frequency scaling impacts the beam transfer function. Figure \ref{fig:sed_wf_bias} shows the ratio of the beam transfer function of the four chromatic beams described above and the one of the nominal case where no frequency scaling was implemented. The chromaticity effect is smooth across all frequency bands and decreases in amplitude with increasing frequency. In the case of the \SI{93}-{GHz} band, not taking account of the beam frequency scaling can result in a transfer function bias as large as $\sim 10 \%$ at $\ell = 700$.

\section{Signal Strength Estimation for different sources}
\label{appendix:snr_calculations}

The total power (in Watts) received by a radio telescope due to an astrophysical source can be expressed as:
\begin{equation}
    P_{\mathrm{received}} = \iint d\Omega d\nu \tau '(\nu ) A_\mathrm{eff} (\nu) B (\theta, \phi, \nu)  S (\theta -\theta_0, \phi - \phi _0, \nu, T) 
    \label{eq:source1}
\end{equation}
where $A_\mathrm{eff} (\nu)$ is the telescope effective area, $B(\theta, \phi, \nu)$ is the frequency-dependent beam response of the telescope, $S (\theta, \phi, \nu, T)$ captures the spectral energy distribution of the source parametrized using the Planck blackbody equation and the thermodynamic temperature $T$,  and $\tau '(\nu)$ captures the spectral response function of the telescope, including effects from the finite transmissivity of the Earth's atmosphere. The planets' thermodynamic temperatures were taken from \cite{planck_planetflux_2017} and we have used $S(\nu, T)$ to represent the Planck blackbody formula for spectral radiance instead of $B(\nu, T)$ to prevent confusion with the beam response.

If the source is small relative to the size of the telescope's beam response subtended on the sky we can collapse the solid angle convolution and write 
\begin{equation}
P_{\mathrm{received}} \approx \frac{\Omega_{\mathrm{source}}}{\Omega_{\mathrm{beam}}}\int S (\nu, T) \tau (\nu) d\nu \approx A_{\mathrm{eff}} \Omega_{\mathrm{source}} \int S (\nu, T) \tau '(\nu) d\nu ,
\label{eq:source2}
\end{equation}
where we have assumed that the spectral response function, $\tau (\nu)$, can be written as
\begin{equation}
\tau (\nu) \equiv \tau ' (\nu) \cdot n \lambda ^2 = \tau ' (\nu) \cdot A_\mathrm{eff} (\nu) \Omega_{\mathrm{beam}} (\nu), 
\label{eq:tau_nu}
\end{equation}
with $n$ corresponding to the number of radiation modes \citep{Hudson_74, Hodara_Slemon_1984}. In Equation \ref{eq:source2}, we have made the approximation that  the frequency dependence of the beam solid angle, $\Omega_{\mathrm{beam}} (\nu)$, or correspondingly the telescope effective area, $A _{\mathrm{eff}} (\nu)$, can be ignored. The accuracy of this approximation depends, of course, on the width of the frequency range over which we must integrate.

For a restricted range of frequencies centered on $\nu = \nu _0$, the above equation can be further simplified to 
\begin{equation}
    P_{\mathrm{received}} \approx A_{\mathrm{eff}} \Omega_{\mathrm{source}}  \cdot S (\nu_0, T) \tau ' (\nu _0 ) \Delta \nu ,
    \label{eq:source3}
\end{equation}
where $\Delta \nu$ corresponds to the frequency bandwidth over which the signal from the source is integrated and $\nu _0$ is the band center frequency. In the above equation, $\tau '(\nu _0)$ describes the telescope's spectral response function that ignores the $n \lambda^2$ scaling (see Equation \ref{eq:tau_nu}). Note, however, that losses in signal strength due to instrumental effects such as non-ideal optical efficiency should be included in $\tau '(\nu _0)$. From Equation \ref{eq:source3}, it is clear that we can calculate the signal strength from a compact source, assuming that we know all of the parameters on the right-hand side of the equation. We use this equation to calculate the signal amplitude from Jupiter, Mars and Saturn. 

The power incident on a telescope from an artificial source has been estimated using the Friis Equation \citep{Friis_1946, Johnson_2015}:
\begin{equation}
    P_{\mathrm{incident}} = \left( \frac{P_{\mathrm{emit}}\cdot A_{\mathrm{tel}}}{4\pi d^2} \right) 10^{g/10}
    \label{eq:drone1}
\end{equation}
where $P_{\mathrm{emit}}$ is the in-band power emitted by the source, $g=10 \log _{10} (4\pi/\Omega _\mathrm{ant})$ is the forward gain of the antenna, and $d$ is the distance from the source. Note that this expression can be extended to explicitly include integration over spectral bandpass, but for simplification, it is common to assume that all power emitted by the source is in the spectral band of the receiver. 

 As in the case of power from a compact astrophysical source, the power arriving on the detector element must account for attenuation due to the optical elements between the outside of the telescope and the focal plane itself. We therefore write:
\begin{equation}
    P_\mathrm{received} = \eta P_\mathrm{incident}
\end{equation}
where $\eta$ describes the optical efficiency of the telescope. The atmosphere is considered transparent for the artificial source ($ \mathcal{T} \simeq 1$) given the simulated frequencies and the short distance of the artificial source, $d \simeq 500 \si{\meter}$.

Finally, the signal-to-noise ratio (SNR) from a compact astrophysical source or a transmitter mounted on a drone is estimated as:
\begin{equation}
    \mathrm{SNR} = \frac{P_{\mathrm{received}}}{\mathrm{NEP}\cdot \sqrt{2/t}},
\end{equation}
where NEP represents the assumed Noise-Equivalent-Power,  and $t$ is the integration time per pixel which we have set to $1 \si{s}$ for generating numbers for Figure \ref{fig:planet_vs_drone}.

\section{Beam transfer function uncertainty as a function of the number of observations}
\label{appendix:bl_f_nobs}

The scaling of the beam fitting performance with the number of input simulations needs to be quantified. From Figure \ref{fig:all_freqs_snr}, we conclude an increase in the SNR, by about half an order of magnitude when increasing the number of available observations by a factor of ten. Ideally, we would like the benefit of increasing the input number of observations to be strongly pronounced on the \texttt{beamlib} results, as well.

\begin{figure}[h!]
\centering
\includegraphics[width=0.8\textwidth]{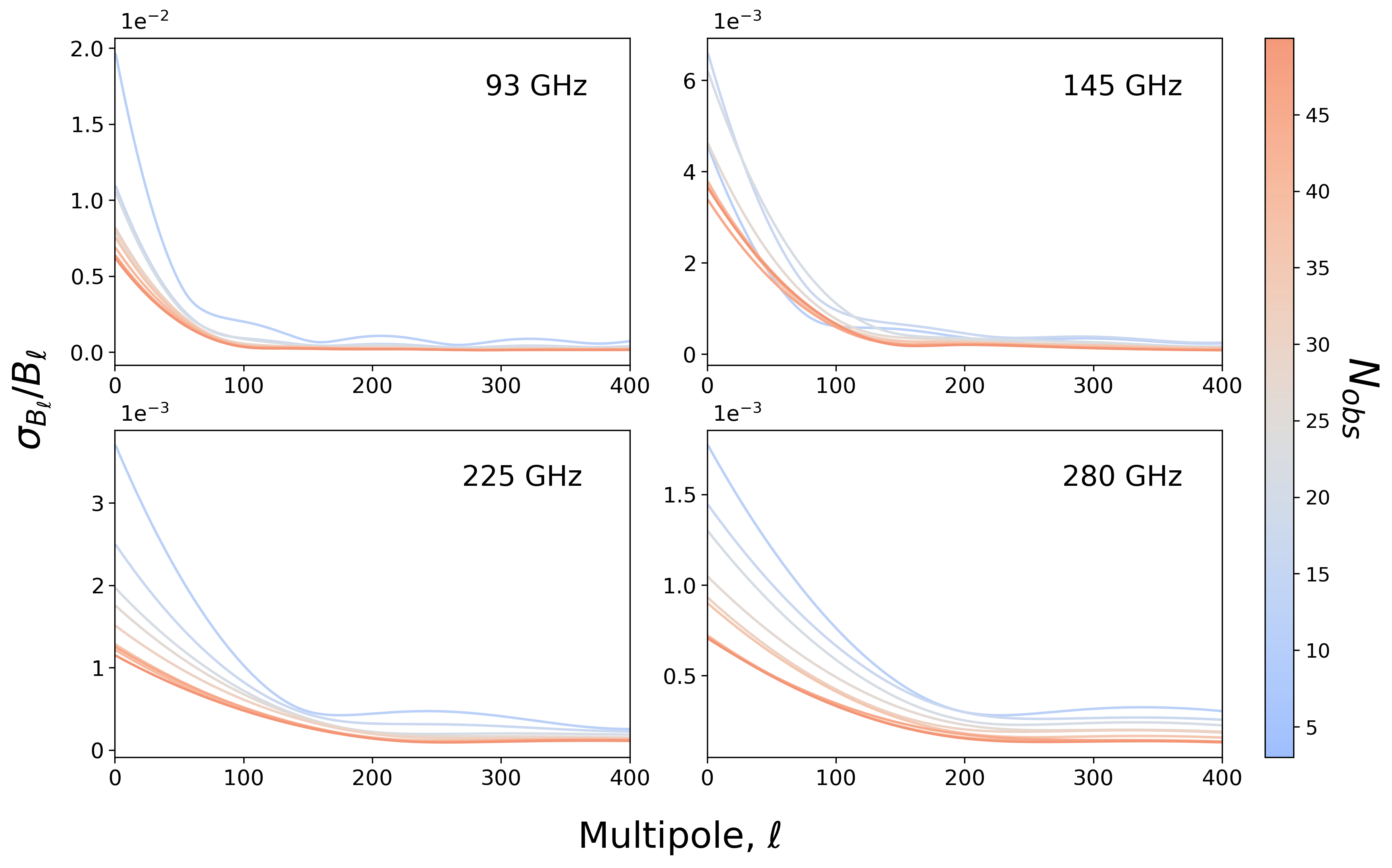}
\caption{\label{fig:all_freqs_luncertainty}Standard deviation of the beam transfer function uncertainty distribution, as calculated from the 10 strongest error modes, divided by the best-fit transfer function value. The standard deviation is plotted versus the number of input observations provided to \texttt{beamlib}, for four frequency bands centred on 93, 145, 225 and \SI{280}{\giga\hertz}.}
\end{figure}

We test this by running the beam fitting code for the same parameters used for the results shown in Figures \ref{fig:all_freqs_profs} and  \ref{fig:all_freqs_wfs_errors} but alter the number of input observations we provide to the code each time. This number ranges again from 5 to 50 observations. We construct distributions of $10^{3}$ samples of the 10 strongest error modes of the recovered beam transfer function, as calculated from \texttt{beamlib}, and we estimate the standard deviation of these distributions per observation number and frequency band. 

Figure \ref{fig:all_freqs_luncertainty} shows the fractional standard deviation of the beam transfer function uncertainty distribution with respect to the best-fit estimation as a function of the observation number for four frequency bands centred on 93, 145, 225 and \SI{280}{\giga\hertz}, respectively. The uncertainty decreases by a factor of $\gtrsim$ 2 when increasing the number of input observations we feed to the code by a factor of ten. This result is consistent for all four frequency bands. 

\bibliography{beam_characterization.bib}{}

\begin{thebibliography}{}
\expandafter\ifx\csname natexlab\endcsname\relax\def\natexlab#1{#1}\fi
\providecommand{\url}[1]{\href{#1}{#1}}
\providecommand{\dodoi}[1]{doi:~\href{http://doi.org/#1}{\nolinkurl{#1}}}
\providecommand{\doeprint}[1]{\href{http://ascl.net/#1}{\nolinkurl{http://ascl.net/#1}}}
\providecommand{\doarXiv}[1]{\href{https://arxiv.org/abs/#1}{\nolinkurl{https://arxiv.org/abs/#1}}}

\bibitem[{Abitbol {et~al.}(2021)Abitbol, Alonso, Simon, Lashner, Crowley, Ali, Azzoni, Baccigalupi, Barron, Brown, \& et~al.}]{Abitbol_2021}
Abitbol, M.~H., Alonso, D., Simon, S.~M., {et~al.} 2021, Journal of Cosmology and Astroparticle Physics, 2021, 032, \dodoi{10.1088/1475-7516/2021/05/032}

\bibitem[{Adler \& Gudmundsson(2020)}]{Adler_2020}
Adler, A., \& Gudmundsson, J.~E. 2020, Millimeter, Submillimeter, and Far-Infrared Detectors and Instrumentation for Astronomy X, \dodoi{10.1117/12.2576309}

\bibitem[{Aikin {et~al.}(2010)Aikin, Ade, Benton, Bock, Bonetti, Brevik, Dowell, Duband, Filippini, Golwala, Halpern, Hristov, Irwin, Kaufman, Keating, Kovac, Kuo, Lange, Netterfield, Nguyen, IV, Orlando, Pryke, Richter, Ruhl, Runyan, Sheehy, Stokes, Sudiwala, Teply, Tolan, Turner, Wilson, \& Wong}]{bicep2_beamcal2010}
Aikin, R.~W., Ade, P.~A., Benton, S., {et~al.} 2010, in Millimeter, Submillimeter, and Far-Infrared Detectors and Instrumentation for Astronomy V, ed. W.~S. Holland \& J.~Zmuidzinas, Vol. 7741, International Society for Optics and Photonics (SPIE), 77410V, \dodoi{10.1117/12.857868}

\bibitem[{Ali {et~al.}(2020)Ali, Adachi, Arnold, Ashton, Bazarko, Chinone, Coppi, Corbett, Crowley, Crowley, \& et~al.}]{Ali_2020}
Ali, A.~M., Adachi, S., Arnold, K., {et~al.} 2020, Journal of Low Temperature Physics, 200, 461–471, \dodoi{10.1007/s10909-020-02430-5}

\bibitem[{Bennett {et~al.}(2013)Bennett, Larson, Weiland, Jarosik, Hinshaw, Odegard, Smith, Hill, Gold, Halpern, Komatsu, Nolta, Page, Spergel, Wollack, Dunkley, Kogut, Limon, Meyer, Tucker, \& Wright}]{Bennett_2013}
Bennett, C.~L., Larson, D., Weiland, J.~L., {et~al.} 2013, The Astrophysical Journal Supplement Series, 208, 20, \dodoi{10.1088/0067-0049/208/2/20}

\bibitem[{{{BICEP}2/Keck Array {XI}} {et~al.}(2019){{BICEP}2/Keck Array {XI}}, Ade, Ahmed, Aikin, Barkats, Benton, Bischoff, Bock, Bowens-Rubin, Brevik, Buder, Bullock, Buza, Connors, Cornelison, Crill, Crumrine, Dierickx, Duband, Filippini, Fliescher, Grayson, Hall, Halpern, Harrison, Hildebrandt, Hilton, Hui, Irwin, Kang, Karkare, Karpel, Kaufman, Keating, Kefeli, Kernasovskiy, Kovac, Kuo, Larsen, Lau, Leitch, Lueker, Megerian, Moncelsi, Namikawa, Netterfield, Nguyen, O'Brient, IV, Palladino, Pryke, Racine, Richter, Schillaci, Schwarz, Sheehy, Soliman, Germaine, Staniszewski, Steinbach, Sudiwala, Teply, Thompson, Tolan, Tucker, Turner, Umilta, Vieregg, Wandui, Weber, Wiebe, Willmert, Wong, Wu, Yang, Yoon, \& and}]{BK2019}
{{BICEP}2/Keck Array {XI}}, Ade, P. A.~R., Ahmed, Z., {et~al.} 2019, The Astrophysical Journal, 884, 114, \dodoi{10.3847/1538-4357/ab391d}

\bibitem[{Choi {et~al.}(2020)Choi, Hasselfield, Ho, Koopman, Lungu, Abitbol, Addison, Ade, Aiola, Alonso, Amiri, Amodeo, Angile, Austermann, Baildon, Battaglia, Beall, Bean, Becker, Bond, Bruno, Calabrese, Calafut, Campusano, Carrero, Chesmore, mei Cho, Clark, Cothard, Crichton, Crowley, Darwish, Datta, Denison, Devlin, Duell, Duff, Duivenvoorden, Dunkley, Dünner, Essinger-Hileman, Fankhanel, Ferraro, Fox, Fuzia, Gallardo, Gluscevic, Golec, Grace, Gralla, Guan, Hall, Halpern, Han, Hargrave, Henderson, Hensley, Hill, Hilton, Hilton, Hincks, Hlo{\v{z}}ek, Hubmayr, Huffenberger, Hughes, Infante, Irwin, Jackson, Klein, Knowles, Kosowsky, Lakey, Li, Li, Li, Lokken, Louis, MacInnis, Madhavacheril, Maldonado, Mallaby-Kay, Marsden, Maurin, McMahon, Menanteau, Moodley, Morton, Naess, Namikawa, Nati, Newburgh, Nibarger, Nicola, Niemack, Nolta, Orlowski-Sherer, Page, Pappas, Partridge, Phakathi, Prince, Puddu, Qu, Rivera, Robertson, Rojas, Salatino, Schaan, Schillaci, Schmitt, Sehgal, Sherwin, Sierra, Sievers, Sifon,
  Sikhosana, Simon, Spergel, Staggs, Stevens, Storer, Sunder, Switzer, Thorne, Thornton, Trac, Treu, Tucker, Vale, Engelen, Lanen, Vavagiakis, Wagoner, Wang, Ward, Wollack, Xu, Zago, \& Zhu}]{Choi_2020}
Choi, S.~K., Hasselfield, M., Ho, S.-P.~P., {et~al.} 2020, Journal of Cosmology and Astroparticle Physics, 2020, 045, \dodoi{10.1088/1475-7516/2020/12/045}

\bibitem[{Coppi {et~al.}(2022)Coppi, Conenna, Savorgnano, Carrero, D\"unner-Planella, Galitzki, Nati, \& Zannoni}]{Coppi:2022qjs}
Coppi, G., Conenna, G., Savorgnano, S., {et~al.} 2022, in {SPIE Astronomical Telescopes + Instrumentation 2022}.
\newblock \doarXiv{2207.07595}

\bibitem[{Duivenvoorden {et~al.}(2021)Duivenvoorden, Adler, Billi, Dachlythra, \& Gudmundsson}]{beamconv_hwp_2021}
Duivenvoorden, A.~J., Adler, A.~E., Billi, M., Dachlythra, N., \& Gudmundsson, J.~E. 2021, Monthly Notices of the Royal Astronomical Society, 502, 4526–4539, \dodoi{10.1093/mnras/stab317}

\bibitem[{{D{\"u}nner} {et~al.}(2020){D{\"u}nner}, {Flux{\'a}}, {Best}, \& {Carrero}}]{2020SPIE11453E..2PD}
{D{\"u}nner}, R., {Flux{\'a}}, J., {Best}, S., \& {Carrero}, F. 2020, in Society of Photo-Optical Instrumentation Engineers (SPIE) Conference Series, Vol. 11453, Society of Photo-Optical Instrumentation Engineers (SPIE) Conference Series, 114532P, \dodoi{10.1117/12.2561165}

\bibitem[{Dünner {et~al.}(2021)Dünner, Fluxá, Best, Carrero, \& Boettger}]{9411058}
Dünner, R., Fluxá, J., Best, S., Carrero, F., \& Boettger, D. 2021, in 2021 15th European Conference on Antennas and Propagation (EuCAP), 1--5, \dodoi{10.23919/EuCAP51087.2021.9411058}

\bibitem[{Dünner {et~al.}(2012)Dünner, Hasselfield, Marriage, Sievers, Acquaviva, Addison, Ade, Aguirre, Amiri, Appel, Barrientos, Battistelli, Bond, Brown, Burger, Calabrese, Chervenak, Das, Devlin, Dicker, Doriese, Dunkley, Essinger-Hileman, Fisher, Gralla, Fowler, Hajian, Halpern, Hern{\'{a} }ndez-Monteagudo, Hilton, Hilton, Hincks, Hlozek, Huffenberger, Hughes, Hughes, Infante, Irwin, Juin, Kaul, Klein, Kosowsky, Lau, Limon, Lin, Louis, Lupton, Marsden, Martocci, Mauskopf, Menanteau, Moodley, Moseley, Netterfield, Niemack, Nolta, Page, Parker, Partridge, Quintana, Reid, Sehgal, Sherwin, Spergel, Staggs, Swetz, Switzer, Thornton, Trac, Tucker, Warne, Wilson, Wollack, \& Zhao}]{D_nner_2012}
Dünner, R., Hasselfield, M., Marriage, T.~A., {et~al.} 2012, The Astrophysical Journal, 762, 10, \dodoi{10.1088/0004-637x/762/1/10}

\bibitem[{Errard {et~al.}(2015)Errard, Ade, Akiba, Arnold, Atlas, Baccigalupi, Barron, Boettger, Borrill, Chapman, Chinone, Cukierman, Delabrouille, Dobbs, Ducout, Elleflot, Fabbian, Feng, Feeney, Gilbert, Goeckner-Wald, Halverson, Hasegawa, Hattori, Hazumi, Hill, Holzapfel, Hori, Inoue, Jaehnig, Jaffe, Jeong, Katayama, Kaufman, Keating, Kermish, Keskitalo, Kisner, Jeune, Lee, Leitch, Leon, Linder, Matsuda, Matsumura, Miller, Myers, Navaroli, Nishino, Okamura, Paar, Peloton, Poletti, Puglisi, Rebeiz, Reichardt, Richards, Ross, Rotermund, Schenck, Sherwin, Siritanasak, Smecher, Stebor, Steinbach, Stompor, Suzuki, Tajima, Takakura, Tikhomirov, Tomaru, Whitehorn, Wilson, Yadav, \& Zahn}]{Errard_2015}
Errard, J., Ade, P. A.~R., Akiba, Y., {et~al.} 2015, The Astrophysical Journal, 809, 63, \dodoi{10.1088/0004-637x/809/1/63}

\bibitem[{Fisher(1922)}]{Fisher_1922saa}
Fisher, R.~A. 1922, Phil. Trans. Roy. Soc. Lond. A, 222, 309, \dodoi{10.1098/rsta.1922.0009}

\bibitem[{Fraisse {et~al.}(2013)Fraisse, Ade, Amiri, Benton, Bock, Bond, Bonetti, Bryan, Burger, Chiang, Clark, Contaldi, Crill, Davis, Dor{\'{e} }, Farhang, Filippini, Fissel, Gandilo, Golwala, Gudmundsson, Hasselfield, Hilton, Holmes, Hristov, Irwin, Jones, Kuo, MacTavish, Mason, Montroy, Morford, Netterfield, O{\textquotesingle}Dea, Rahlin, Reintsema, Ruhl, Runyan, Schenker, Shariff, Soler, Trangsrud, Tucker, Tucker, Turner, \& Wiebe}]{Fraisse_2013}
Fraisse, A., Ade, P., Amiri, M., {et~al.} 2013, Journal of Cosmology and Astroparticle Physics, 2013, 047, \dodoi{10.1088/1475-7516/2013/04/047}

\bibitem[{Friis(1946)}]{Friis_1946}
Friis, H. 1946, Proceedings of the IRE, 34, 254, \dodoi{10.1109/JRPROC.1946.234568}

\bibitem[{Galitzki(2018)}]{galitzki2018simons}
Galitzki, N. 2018, The Simons Observatory: Project Overview.
\newblock \doarXiv{1810.02465}

\bibitem[{{Global Modeling and Assimilation Office (GMAO)}((2015))}]{merra2}
{Global Modeling and Assimilation Office (GMAO)}. (2015), MERRA-2 tavg1\_2d\_slv\_Nx: 2d,1-Hourly,Time-Averaged,Single-Level,Assimilation,Single-Level Diagnostics V5.12.4, https://gmao.gsfc.nasa.gov/reanalysis/MERRA-2/citing\_MERRA-2/

\bibitem[{Hasselfield {et~al.}(2013)Hasselfield, Moodley, Bond, Das, Devlin, Dunkley, Dünner, Fowler, Gallardo, Gralla, \& et~al.}]{Hasselfield_2013}
Hasselfield, M., Moodley, K., Bond, J.~R., {et~al.} 2013, The Astrophysical Journal Supplement Series, 209, 17, \dodoi{10.1088/0067-0049/209/1/17}

\bibitem[{Hill {et~al.}(2018)Hill, Kusaka, Barton, Bixler, Droster, Flament, Ganjam, Jadbabaie, Jeong, Lee, \& et~al.}]{Hill_2018}
Hill, C.~A., Kusaka, A., Barton, P., {et~al.} 2018, Journal of Low Temperature Physics, 193, 851–859, \dodoi{10.1007/s10909-018-1980-6}

\bibitem[{Hodara \& Slemon(1984)}]{Hodara_Slemon_1984}
Hodara, H., \& Slemon, C. 1984, Flow Turbulence and Combustion - FLOW TURBUL COMBUST, 41, 203, \dodoi{10.1007/BF00382453}

\bibitem[{Hudson(1974)}]{Hudson_74}
Hudson, M.~C. 1974, Appl. Opt., 13, 1029, \dodoi{10.1364/AO.13.001029}

\bibitem[{{Johnson} {et~al.}(2015){Johnson}, {Vourch}, {Drysdale}, {Kalman}, {Fujikawa}, {Keating}, \& {Kaufman}}]{Johnson2015}
{Johnson}, B.~R., {Vourch}, C.~J., {Drysdale}, T.~D., {et~al.} 2015, Journal of Astronomical Instrumentation, 4, 1550007, \dodoi{10.1142/S2251171715500075}

\bibitem[{Johnson(1992)}]{Johnson_2015}
Johnson, R.~C. 1992, Antenna Engineering Handbook, 3rd edition (McGraw-Hill Professional)

\bibitem[{Keating {et~al.}(2012)Keating, Shimon, \& Yadav}]{Keating_2013}
Keating, B.~G., Shimon, M., \& Yadav, A. P.~S. 2012, The Astrophysical Journal Letters, 762, L23, \dodoi{10.1088/2041-8205/762/2/L23}

\bibitem[{Kolmogorov(1941)}]{Kolmogorov}
Kolmogorov, A. 1941, Doklady Akademii Nauk SSSR, 30, 301-304.

\bibitem[{{Komatsu}(2022)}]{new_physics_from_CMBpol}
{Komatsu}, E. 2022, arXiv e-prints, arXiv:2202.13919.
\newblock \doarXiv{2202.13919}

\bibitem[{Krachmalnicoff {et~al.}(2016)Krachmalnicoff, Baccigalupi, Aumont, Bersanelli, \& Mennella}]{Krachmalnicoff_2016}
Krachmalnicoff, N., Baccigalupi, C., Aumont, J., Bersanelli, M., \& Mennella, A. 2016, Astronomy \& Astrophysics, 588, A65, \dodoi{10.1051/0004-6361/201527678}

\bibitem[{Lungu {et~al.}(2022)Lungu, Storer, Hasselfield, Duivenvoorden, Calabrese, Chesmore, Choi, Dunkley, Dünner, Gallardo, Golec, Guan, Hill, Hincks, Hubmayr, Madhavacheril, Mallaby-Kay, McMahon, Moodley, Naess, Nati, Niemack, Page, Partridge, Puddu, Schillaci, Sif{\'{o}}n, Staggs, Sunder, Wollack, \& Xu}]{Lungu_2022}
Lungu, M., Storer, E.~R., Hasselfield, M., {et~al.} 2022, Journal of Cosmology and Astroparticle Physics, 2022, 044, \dodoi{10.1088/1475-7516/2022/05/044}

\bibitem[{Masi {et~al.}(2006)Masi, Ade, Bock, Bond, Borrill, Boscaleri, Cabella, Contaldi, Crill, de~Bernardis, Gasperis, de~Oliveira-Costa, Troia, Stefano, Ehlers, Hivon, Hristov, Iacoangeli, Jaffe, Jones, Kisner, Lange, MacTavish, Bettolo, Mason, Mauskopf, Montroy, Nati, Nati, Natoli, Netterfield, Pascale, Piacentini, Pogosyan, Polenta, Prunet, Ricciardi, Romeo, Ruhl, Santini, Tegmark, Torbet, Veneziani, \& Vittorio}]{Masi_2006}
Masi, S., Ade, P. A.~R., Bock, J.~J., {et~al.} 2006, Astronomy \& Astrophysics, 458, 687, \dodoi{10.1051/0004-6361:20053891}

\bibitem[{Matsuda(2020)}]{sat_optics_design}
Matsuda, F.~T. 2020, Ground-based and Airborne Telescopes VIII, \dodoi{10.1117/12.2561205}

\bibitem[{Morris {et~al.}(2022)Morris, Bustos, Calabrese, Choi, Duivenvoorden, Dunkley, Dünner, Gallardo, Hasselfield, Hincks, Mroczkowski, Naess, Niemack, Page, Partridge, Salatino, Staggs, Treu, Wollack, \& Xu}]{Morris_2022}
Morris, T.~W., Bustos, R., Calabrese, E., {et~al.} 2022, Physical Review D, 105, \dodoi{10.1103/physrevd.105.042004}

\bibitem[{N{\ae}ss(2019)}]{Naess_2019_x}
N{\ae}ss, S.~K. 2019, Journal of Cosmology and Astroparticle Physics, 2019, 060, \dodoi{10.1088/1475-7516/2019/12/060}

\bibitem[{Nati {et~al.}(2017)Nati, Devlin, Gerbino, Johnson, Keating, Pagano, \& Teply}]{Nati_2017}
Nati, F., Devlin, M.~J., Gerbino, M., {et~al.} 2017, Journal of Astronomical Instrumentation, 06, \dodoi{10.1142/s2251171717400086}

\bibitem[{Puglisi {et~al.}(2021)Puglisi, Keskitalo, Kisner, \& Borrill}]{TOAST_systematics_2021}
Puglisi, G., Keskitalo, R., Kisner, T., \& Borrill, J.~D. 2021, Research Notes of the AAS, 5, 137, \dodoi{10.3847/2515-5172/ac0823}

\bibitem[{Salatino {et~al.}(2018)Salatino, Lashner, Gerbino, Simon, Didier, Ali, Ashton, Bryan, Chinone, Coughlin, Crowley, Fabbian, Galitzki, Goeckner-Wald, Golec, Gudmundsson, Hill, Keating, Kusaka, Lee, McMahon, Miller, Puglisi, Reichardt, Teply, Xu, \& Zhu}]{salatino2018studies}
Salatino, M., Lashner, J., Gerbino, M., {et~al.} 2018, Studies of Systematic Uncertainties for Simons Observatory: Polarization Modulator Related Effects.
\newblock \doarXiv{1808.07442}

\bibitem[{Shimon {et~al.}(2008)Shimon, Keating, Ponthieu, \& Hivon}]{Shimon_2008}
Shimon, M., Keating, B., Ponthieu, N., \& Hivon, E. 2008, Phys. Rev. D, 77, 083003, \dodoi{10.1103/PhysRevD.77.083003}

\bibitem[{Takakura {et~al.}(2019)Takakura, Aguilar-Faúndez, Akiba, Arnold, Baccigalupi, Barron, Beck, Bianchini, Boettger, Borrill, \& et~al.}]{atm_pol_2019}
Takakura, S., Aguilar-Faúndez, M. A.~O., Akiba, Y., {et~al.} 2019, The Astrophysical Journal, 870, 102, \dodoi{10.3847/1538-4357/aaf381}

\bibitem[{{The Planck Collaboration} {et~al.}(2020){The Planck Collaboration}, Akrami, Ashdown, Aumont, Baccigalupi, Ballardini, Banday, Barreiro, Bartolo, Basak, \& et~al.}]{planck_diff_sep_comp_2020}
{The Planck Collaboration}, Akrami, Y., Ashdown, M., {et~al.} 2020, Astronomy \& Astrophysics, 641, A4, \dodoi{10.1051/0004-6361/201833881}

\bibitem[{{The Planck Collaboration I} {et~al.}(2020){The Planck Collaboration I}, Aghanim, Akrami, Arroja, Ashdown, Aumont, Baccigalupi, Ballardini, Banday, Barreiro, \& et~al.}]{planck_overview_2020}
{The Planck Collaboration I}, Aghanim, N., Akrami, Y., {et~al.} 2020, Astronomy \& Astrophysics, 641, A1, \dodoi{10.1051/0004-6361/201833880}

\bibitem[{{The Planck Collaboration LII} {et~al.}(2017){The Planck Collaboration LII}, Akrami, Ashdown, Aumont, Baccigalupi, Ballardini, Banday, Barreiro, Bartolo, Basak, \& et~al.}]{planck_planetflux_2017}
{The Planck Collaboration LII}, Akrami, Y., Ashdown, M., {et~al.} 2017, Astronomy \& Astrophysics, 607, A122, \dodoi{10.1051/0004-6361/201630311}

\bibitem[{{The Planck Collaboration VII} {et~al.}(2016){The Planck Collaboration VII}, {Adam, R.}, {Ade, P. A. R.}, {Aghanim, N.}, {Arnaud, M.}, {Ashdown, M.}, {Aumont, J.}, {Baccigalupi, C.}, {Banday, A. J.}, {Barreiro, R. B.}, {Bartolo, N.}, {Battaner, E.}, {Benabed, K.}, {Beno\^{\i}t, A.}, {Benoit-L\'evy, A.}, {Bernard, J.-P.}, {Bersanelli, M.}, {Bertincourt, B.}, {Bielewicz, P.}, {Bock, J. J.}, {Bonavera, L.}, {Bond, J. R.}, {Borrill, J.}, {Bouchet, F. R.}, {Boulanger, F.}, {Bucher, M.}, {Burigana, C.}, {Calabrese, E.}, {Cardoso, J.-F.}, {Catalano, A.}, {Challinor, A.}, {Chamballu, A.}, {Chary, R.-R.}, {Chiang, H. C.}, {Christensen, P. R.}, {Clements, D. L.}, {Colombi, S.}, {Colombo, L. P. L.}, {Combet, C.}, {Couchot, F.}, {Coulais, A.}, {Crill, B. P.}, {Curto, A.}, {Cuttaia, F.}, {Danese, L.}, {Davies, R. D.}, {Davis, R. J.}, {de Bernardis, P.}, {de Rosa, A.}, {de Zotti, G.}, {Delabrouille, J.}, {Delouis, J.-M.}, {D\'esert, F.-X.}, {Diego, J. M.}, {Dole, H.}, {Donzelli, S.}, {Dor\'e, O.}, {Douspis, M.},
  {Ducout, A.}, {Dupac, X.}, {Efstathiou, G.}, {Elsner, F.}, {En\ss{}lin, T. A.}, {Eriksen, H. K.}, {Falgarone, E.}, {Fergusson, J.}, {Finelli, F.}, {Forni, O.}, {Frailis, M.}, {Fraisse, A. A.}, {Franceschi, E.}, {Frejsel, A.}, {Galeotta, S.}, {Galli, S.}, {Ganga, K.}, {Ghosh, T.}, {Giard, M.}, {Giraud-H\'eraud, Y.}, {Gjerl\o{}w, E.}, {Gonz\'alez-Nuevo, J.}, {G\'orski, K. M.}, {Gratton, S.}, {Gruppuso, A.}, {Gudmundsson, J. E.}, {Hansen, F. K.}, {Hanson, D.}, {Harrison, D. L.}, {Henrot-Versill\'e, S.}, {Herranz, D.}, {Hildebrandt, S. R.}, {Hivon, E.}, {Hobson, M.}, {Holmes, W. A.}, {Hornstrup, A.}, {Hovest, W.}, {Huffenberger, K. M.}, {Hurier, G.}, {Jaffe, A. H.}, {Jaffe, T. R.}, {Jones, W. C.}, {Juvela, M.}, {Keih\"anen, E.}, {Keskitalo, R.}, {Kisner, T. S.}, {Kneissl, R.}, {Knoche, J.}, {Kunz, M.}, {Kurki-Suonio, H.}, {Lagache, G.}, {Lamarre, J.-M.}, {Lasenby, A.}, {Lattanzi, M.}, {Lawrence, C. R.}, {Le Jeune, M.}, {Leahy, J. P.}, {Lellouch, E.}, {Leonardi, R.}, {Lesgourgues, J.}, {Levrier, F.}, {Liguori,
  M.}, {Lilje, P. B.}, {Linden-V\o{}rnle, M.}, {L\'opez-Caniego, M.}, {Lubin, P. M.}, {Mac\'{\i}as-P\'erez, J. F.}, {Maggio, G.}, {Maino, D.}, {Mandolesi, N.}, {Mangilli, A.}, {Maris, M.}, {Martin, P. G.}, {Mart\'{\i}nez-Gonz\'alez, E.}, {Masi, S.}, {Matarrese, S.}, {McGehee, P.}, {Melchiorri, A.}, {Mendes, L.}, {Mennella, A.}, {Migliaccio, M.}, {Mitra, S.}, {Miville-Desch\^enes, M.-A.}, {Moneti, A.}, {Montier, L.}, {Moreno, R.}, {Morgante, G.}, {Mortlock, D.}, {Moss, A.}, {Mottet, S.}, {Munshi, D.}, {Murphy, J. A.}, {Naselsky, P.}, {Nati, F.}, {Natoli, P.}, {Netterfield, C. B.}, {N\o{}rgaard-Nielsen, H. U.}, {Noviello, F.}, {Novikov, D.}, {Novikov, I.}, {Oxborrow, C. A.}, {Paci, F.}, {Pagano, L.}, {Pajot, F.}, {Paoletti, D.}, {Pasian, F.}, {Patanchon, G.}, {Pearson, T. J.}, {Perdereau, O.}, {Perotto, L.}, {Perrotta, F.}, {Pettorino, V.}, {Piacentini, F.}, {Piat, M.}, {Pierpaoli, E.}, {Pietrobon, D.}, {Plaszczynski, S.}, {Pointecouteau, E.}, {Polenta, G.}, {Pratt, G. W.}, {Pr\'ezeau, G.}, {Prunet, S.},
  {Puget, J.-L.}, {Rachen, J. P.}, {Reinecke, M.}, {Remazeilles, M.}, {Renault, C.}, {Renzi, A.}, {Ristorcelli, I.}, {Rocha, G.}, {Rosset, C.}, {Rossetti, M.}, {Roudier, G.}, {Rowan-Robinson, M.}, {Rusholme, B.}, {Sandri, M.}, {Santos, D.}, {Sauv\'e, A.}, {Savelainen, M.}, {Savini, G.}, {Scott, D.}, {Seiffert, M. D.}, {Shellard, E. P. S.}, {Spencer, L. D.}, {Stolyarov, V.}, {Stompor, R.}, {Sudiwala, R.}, {Sutton, D.}, {Suur-Uski, A.-S.}, {Sygnet, J.-F.}, {Tauber, J. A.}, {Terenzi, L.}, {Toffolatti, L.}, {Tomasi, M.}, {Tristram, M.}, {Tucci, M.}, {Tuovinen, J.}, {Valenziano, L.}, {Valiviita, J.}, {Van Tent, B.}, {Vibert, L.}, {Vielva, P.}, {Villa, F.}, {Wade, L. A.}, {Wandelt, B. D.}, {Watson, R.}, {Wehus, I. K.}, {Yvon, D.}, {Zacchei, A.}, \& {Zonca, A.}}]{planck_toi_and_beams_2015}
{The Planck Collaboration VII}, {Adam, R.}, {Ade, P. A. R.}, {et~al.} 2016, A\&A, 594, A7, \dodoi{10.1051/0004-6361/201525844}

\bibitem[{{The Simons Observatory} {et~al.}(2019){The Simons Observatory}, {Ade}, {Aguirre}, {Ahmed}, {Aiola}, {Ali}, {Alonso}, {Alvarez}, {Arnold}, {Ashton}, {Austermann}, {Awan}, {Baccigalupi}, {Baildon}, {Barron}, {Battaglia}, {Battye}, {Baxter}, {Bazarko}, {Beall}, {Bean}, {Beck}, {Beckman}, {Beringue}, {Bianchini}, {Boada}, {Boettger}, {Bond}, {Borrill}, {Brown}, {Bruno}, {Bryan}, {Calabrese}, {Calafut}, {Calisse}, {Carron}, {Challinor}, {Chesmore}, {Chinone}, {Chluba}, {Cho}, {Choi}, {Coppi}, {Cothard}, {Coughlin}, {Crichton}, {Crowley}, {Crowley}, {Cukierman}, {D'Ewart}, {D{\"u}nner}, {de Haan}, {Devlin}, {Dicker}, {Didier}, {Dobbs}, {Dober}, {Duell}, {Duff}, {Duivenvoorden}, {Dunkley}, {Dusatko}, {Errard}, {Fabbian}, {Feeney}, {Ferraro}, {Flux{\`a}}, {Freese}, {Frisch}, {Frolov}, {Fuller}, {Fuzia}, {Galitzki}, {Gallardo}, {Tomas Galvez Ghersi}, {Gao}, {Gawiser}, {Gerbino}, {Gluscevic}, {Goeckner-Wald}, {Golec}, {Gordon}, {Gralla}, {Green}, {Grigorian}, {Groh}, {Groppi}, {Guan}, {Gudmundsson}, {Han},
  {Hargrave}, {Hasegawa}, {Hasselfield}, {Hattori}, {Haynes}, {Hazumi}, {He}, {Healy}, {Henderson}, {Hervias-Caimapo}, {Hill}, {Hill}, {Hilton}, {Hilton}, {Hincks}, {Hinshaw}, {Hlo{\v{z}}ek}, {Ho}, {Ho}, {Howe}, {Huang}, {Hubmayr}, {Huffenberger}, {Hughes}, {Ijjas}, {Ikape}, {Irwin}, {Jaffe}, {Jain}, {Jeong}, {Kaneko}, {Karpel}, {Katayama}, {Keating}, {Kernasovskiy}, {Keskitalo}, {Kisner}, {Kiuchi}, {Klein}, {Knowles}, {Koopman}, {Kosowsky}, {Krachmalnicoff}, {Kuenstner}, {Kuo}, {Kusaka}, {Lashner}, {Lee}, {Lee}, {Leon}, {Leung}, {Lewis}, {Li}, {Li}, {Limon}, {Linder}, {Lopez-Caraballo}, {Louis}, {Lowry}, {Lungu}, {Madhavacheril}, {Mak}, {Maldonado}, {Mani}, {Mates}, {Matsuda}, {Maurin}, {Mauskopf}, {May}, {McCallum}, {McKenney}, {McMahon}, {Meerburg}, {Meyers}, {Miller}, {Mirmelstein}, {Moodley}, {Munchmeyer}, {Munson}, {Naess}, {Nati}, {Navaroli}, {Newburgh}, {Nguyen}, {Niemack}, {Nishino}, {Orlowski-Scherer}, {Page}, {Partridge}, {Peloton}, {Perrotta}, {Piccirillo}, {Pisano}, {Poletti}, {Puddu}, {Puglisi},
  {Raum}, {Reichardt}, {Remazeilles}, {Rephaeli}, {Riechers}, {Rojas}, {Roy}, {Sadeh}, {Sakurai}, {Salatino}, {Sathyanarayana Rao}, {Schaan}, {Schmittfull}, {Sehgal}, {Seibert}, {Seljak}, {Sherwin}, {Shimon}, {Sierra}, {Sievers}, {Sikhosana}, {Silva-Feaver}, {Simon}, {Sinclair}, {Siritanasak}, {Smith}, {Smith}, {Spergel}, {Staggs}, {Stein}, {Stevens}, {Stompor}, {Suzuki}, {Tajima}, {Takakura}, {Teply}, {Thomas}, {Thorne}, {Thornton}, {Trac}, {Tsai}, {Tucker}, {Ullom}, {Vagnozzi}, {van Engelen}, {Van Lanen}, {Van Winkle}, {Vavagiakis}, {Verg{\`e}s}, {Vissers}, {Wagoner}, {Walker}, {Ward}, {Westbrook}, {Whitehorn}, {Williams}, {Williams}, {Wollack}, {Xu}, {Yu}, {Yu}, {Zago}, {Zhang}, {Zhu}, \& {Simons Observatory Collaboration}}]{Ade2019}
{The Simons Observatory}, {Ade}, P., {Aguirre}, J., {et~al.} 2019, \jcap, 2019, 056, \dodoi{10.1088/1475-7516/2019/02/056}

\bibitem[{Weiland {et~al.}(2011)Weiland, Odegard, Hill, Wollack, Hinshaw, Greason, Jarosik, Page, Bennett, Dunkley, \& et~al.}]{calibration_wmap_2011}
Weiland, J.~L., Odegard, N., Hill, R.~S., {et~al.} 2011, The Astrophysical Journal Supplement Series, 192, 19, \dodoi{10.1088/0067-0049/192/2/19}

\bibitem[{Wolz {et~al.}(2023)Wolz, Azzoni, Hervias-Caimapo, Errard, Krachmalnicoff, Alonso, Baccigalupi, Lizancos, Brown, Calabrese, Chluba, Dunkley, Fabbian, Galitzki, Jost, Morshed, \& Nati}]{wolz2023simons}
Wolz, K., Azzoni, S., Hervias-Caimapo, C., {et~al.} 2023.
\newblock \doarXiv{2302.04276}

\bibitem[{Xu {et~al.}(2020)Xu, Bhandarkar, Coppi, Kofman, Orlowski-Scherer, Zhu, Ali, Arnold, Austermann, Choi, \& et~al.}]{Xu_2020}
Xu, Z., Bhandarkar, T., Coppi, G., {et~al.} 2020, Millimeter, Submillimeter, and Far-Infrared Detectors and Instrumentation for Astronomy X, \dodoi{10.1117/12.2576151}

\end{thebibliography}
\bibliographystyle{aasjournal}

\end{document}